\documentclass[%
 reprint,
 amsmath,amssymb,
 aps,
]{revtex4-2}

\usepackage{graphicx}
\usepackage{bm}
\usepackage[normalem]{ulem}
\usepackage{color,soul}
\usepackage[colorlinks=true, citecolor=black, urlcolor=blue]{hyperref}
\usepackage{tabularx}
\usepackage{array}
\usepackage{adjustbox}
\usepackage[table,xcdraw]{xcolor}
\usepackage{textcomp}
\usepackage{siunitx}
\usepackage{subfig}
\usepackage{tikz}
\usetikzlibrary{shapes}
\usepackage{multirow}
\usepackage{booktabs}
\usepackage{dcolumn}

\begin{document}

\newcommand{\markersquare}{\raisebox{0.5pt}{\tikz{\node[draw,scale=0.4,regular polygon, regular polygon sides=4,fill=none](){};}}}
\newcommand{\markerdiamond}{\raisebox{0.5pt}{\tikz{\node[draw,scale=0.4,diamond,fill=none](){};}}}
\newcommand{\markerround}{\raisebox{0.5pt}{\tikz{\node[draw,scale=0.4,circle,fill=none](){};}}}
\newcommand{\markerone}{\raisebox{0.5pt}{\tikz{\node[draw,scale=0.4,circle,fill=black](){};}}}

\title{Symbolic Ensemble Learning Enables Discovery of Fast Accurate Physics-Based Interatomic Potentials}

\author{Bilvin Varughese}
\affiliation{Department of Mechanical and Industrial Engineering, University of Illinois, Chicago, Illinois 60607, United States}
\affiliation{Center for Nanoscale Materials, Argonne National Laboratory, Lemont, Illinois 60439, United States}

\author{Aditya Koneru}
\affiliation{Department of Mechanical and Industrial Engineering, University of Illinois, Chicago, Illinois 60607, United States}
\affiliation{Center for Nanoscale Materials, Argonne National Laboratory, Lemont, Illinois 60439, United States}

\author{Adil Muhammad}
\affiliation{Department of Mechanical and Industrial Engineering, University of Illinois, Chicago, Illinois 60607, United States}
\affiliation{Center for Nanoscale Materials, Argonne National Laboratory, Lemont, Illinois 60439, United States}

\author{Troy D. Loeffler}
\affiliation{Department of Mechanical and Industrial Engineering, University of Illinois, Chicago, Illinois 60607, United States}
\affiliation{Center for Nanoscale Materials, Argonne National Laboratory, Lemont, Illinois 60439, United States}

\author{Sukriti Manna}
\affiliation{Department of Mechanical and Industrial Engineering, University of Illinois, Chicago, Illinois 60607, United States}
\affiliation{Center for Nanoscale Materials, Argonne National Laboratory, Lemont, Illinois 60439, United States}

\author{Jan Michael Y. Carrillo}
\affiliation{Center for Nanophase Material Sciences, Oak Ridge National Laboratory, Oak Ridge, TN 37830, United States}

\author{Orcun Yildiz}
\affiliation{Mathematics and Computer Science Division, Argonne National Laboratory, Lemont, Illinois 60439, United States}

\author{Thomas Peterka}
\affiliation{Mathematics and Computer Science Division, Argonne National Laboratory, Lemont, Illinois 60439, United States}

\author{Subramanian K.R.S. Sankaranarayanan}
\email{ssankaranarayanan@anl.gov}
\altaffiliation{skrssank@uic.edu}
\affiliation{Department of Mechanical and Industrial Engineering, University of Illinois, Chicago, Illinois 60607, United States}
\affiliation{Center for Nanoscale Materials, Argonne National Laboratory, Lemont, Illinois 60439, United States}

\date{\today}

\begin{abstract}

Machine learning has transformed materials simulation by delivering force fields with ab initio accuracy, yet bridging the gap between high-dimensional regression and physical interpretability remains a grand challenge. Conventional analytical potentials offer transparency but often fail to capture the complexity of far-from-ground state regimes. Here, we introduce a hybrid symbolic–neural framework that unifies the interpretability of the Embedded Atom Method (EAM) with the adaptability of data-driven learning. Using Equation Learner Neural Networks (EqNNs) trained on density functional theory (DFT) data, we obtain interpretable models for aluminum through three distinct training protocols: random initialization trained via Monte Carlo Tree Search (MCTS) and gradient descent, and two transfer-learning strategies initialized from a copper potential—one employing MCTS followed by gradient descent, and the other using gradient descent only. We find that while all three resulting symbolic models achieve sub-10 meV/atom accuracy, they occupy distinct local minima in the functional landscape, exhibiting complementary trade-offs across phonon dispersion, surface energetics, and elastic response. By integrating these diverse functional forms through a weighted symbolic ensemble, we derive a composite potential that surpasses the fidelity of its constituent models. The resulting ensemble effectively mitigates individual biases, delivering superior consistency with DFT benchmarks across equation-of-state curvature, phonon spectra, and melting dynamics. This approach demonstrates that combining transfer learning with ensemble symbolic regression yields compact, transparent potentials capable of robust prediction across equilibrium and non-equilibrium states.

\end{abstract}

\maketitle

\section{Introduction}

Interatomic potentials form the computational backbone of molecular dynamics simulations, enabling the study of materials phenomena across scales ranging from atomic vibrations to defect migration and plastic deformation. Their accuracy determines the reliability of predicted thermomechanical, electronic, and structural properties while their transferability ensures the model generalizes to atomic environments beyond the training set; and their interpretability depend on having a transparent mathematical form that elucidates the governing physics. Despite decades of development, achieving all three objectives—accuracy, transferability, and interpretability—within a single framework remains one of the grand challenges in computational materials science.\cite{deringer2019machine,jacobs2025practical}. Conventional analytical models, such as the Stillinger-Weber, Tersoff, Embedded Atom Method (EAM), Sutton–Chen, and Finnis–Sinclair potentials \cite{daw1984embedded,sutton1990long,finnis1984simple, stillinger1985computer, backman2012bond, manna2022learning, koneru2022multi}, are founded on physically motivated functional forms and remain indispensable for large-scale simulations owing to their low computational cost. However, their parameterized simplicity restricts their validity to near-equilibrium configurations, limiting their ability to capture defect formation, amorphization, or high-strain behavior. As a result, these traditional approaches struggle to generalize across the complex, multi-regime energy landscapes that modern materials problems demand. Bridging this gap between physical interpretability and non-equilibrium fidelity remains a central challenge in computational materials modeling.

In parallel, machine-learning (ML) force field models \cite{zuo2020performance,nyshadham2019machine,artrith2017efficient,huan2017universal},  have transformed the field by achieving near–density functional theory (DFT) accuracy across diverse configurational and chemical spaces. Models based on neural networks , Gaussian process regression , and graph message passing architectures such as NequIP and MACE  have demonstrated exceptional flexibility in representing high-dimensional potential energy surfaces. Frameworks such as Neural Network Potentials (NNPs) \cite{behler2007generalized,bartok2010gaussian,thompson2015spectral,varughese2024active}, Gaussian Approximation Potentials (GAPs)\cite{bartok2013representing,bartok2015g,banik2024development}, Moment Tensor Potentials (MTPs) \cite{shapeev2016moment}, Spectral Neighbor Analysis Potentials (SNAP)\cite{thompson2015spectral}, and Graph Neural Network (GNN)-based models like NequIP and MACE \cite{batzner20223,batatia2022mace} can achieve near-DFT accuracy across chemically and structurally diverse datasets. These models excel in interpolating complex potential energy surfaces by encoding many-body correlations directly from data, and they have enabled simulations of unprecedented scale and fidelity\cite{bartok2018machine,deringer2018realistic}. Nevertheless, their black-box nature and high computational overhead pose serious drawbacks. Recent advances in machine learning interatomic potentials have introduced formally complete frameworks such as Moment Tensor Potentials (MTP)~\cite{shapeev2016moment} and the Atomic Cluster Expansion (ACE)~\cite{drautz2019atomic,ortner2023atomic}, which provide systematic and in-principle complete basis sets for representing atomic interactions. Despite their mathematical generality, the resulting models typically contain large numbers of expansion coefficients, limiting interpretability and making it difficult to extract clear physical trends or assess robustness outside the training domain. Moreover, such models can require tens of thousands of DFT-labeled configurations and extensive hyperparameter tuning to reach convergence \cite{podryabinkin2017active,jinnouchi2019fly,vandermause2020fly}. The lack of explicit functional forms also complicates embedding these models into existing simulation engines, hinders physical interpretability, and makes it challenging to audit or modify them for new systems or boundary conditions. Consequently, while they rival DFT in accuracy, they fall short in transparency, auditability, and cross-regime robustness.

A complementary strategy gaining traction, especially for developing fast interpretable and accurate potentials, is symbolic regression (SR) - the discovery of explicit mathematical expressions that best fit data without predefined functional forms \cite{schmidt2009distilling,udrescu2020ai,ghiringhelli2015big,hernandez2019fast,wang2024exploring}. SR offers the promise of interpretability and compactness, yielding closed-form equations that can generalize across regimes with modest data requirements. In the context of interatomic potentials, symbolic regression offers the appealing possibility of uncovering interpretable functional forms that retain the expressiveness of machine learning without sacrificing physical meaning\cite{guo2022improving,zhang2023optimizing,tan2022discovery}. Several recent works have applied symbolic regression or related sparse-learning techniques to rediscover known analytic potentials or propose generalized forms for pairwise or many-body interactions. These include approaches such as AI Feynman, PySR, and Operon, which employ evolutionary algorithms or genetic programming to search for compact mathematical expressions that fit atomic-scale data \cite{udrescu2020ai, cranmer2023interpretable,burlacu2020operon}. While promising, most symbolic regression frameworks to date rely on purely stochastic search strategies, which can be slow to converge, prone to overfitting, or limited to small datasets. They also typically treat symbolic regression as an equation-fitting problem rather than a learned hierarchical process, leading to isolated equations rather than structured, physically consistent potential frameworks. Furthermore, they seldom incorporate physical constraints—such as positivity of pair and density terms or domain-specific penalties—which are critical for ensuring meaningful and stable functional forms.

In this work, we introduce a hybrid symbolic–neural framework that unifies the interpretability of symbolic regression with the adaptability of machine learning. Our approach employs Equation Neural Networks (EqNNs)\cite{martius2016extrapolation,sahoo2018learning,werner2021informed,chen2020learning}—a class of neural architectures in which standard activations are replaced by functional activations corresponding to elementary mathematical operations. When trained on atomistic datasets, EqNNs collapse into compact symbolic equations that can be directly interpreted as analytic functional forms. We implement this framework within the Embedded Atom Method (EAM) formalism\cite{sutton1990long}, constructing three coupled neural subnetworks that represent the pair, density, and embedding contributions to the total energy. This architecture ensures physical consistency while allowing the discovery of novel, data-driven equations that extend beyond empirical EAM forms. The training objective incorporates physics-aware constraints, including hard penalties for non-physical outputs (e.g., negative pair or density values) and domain-related instabilities (NaNs or infinities), ensuring that the learned symbolic forms remain both stable and physically meaningful.

 A distinguishing aspect of our study is the deliberate inclusion of highly distorted, far-from-ground-state configurations during training. We define these configurations as structures whose DFT energy per atom exceeds the DFT ground-state reference by more than 0.5 eV/atom. This cutoff, approximately 15\% of our calculated aluminum cohesive energy, provides an operational division of the dataset rather than a thermodynamic boundary. Although strongly distorted configurations are rarely sampled under near-equilibrium conditions, they probe interactions beyond the low-energy region, including short-range repulsion. Including them provides additional constraints on the learned potential and a more complete representation of the potential-energy landscape.  Using a nested ensemble sampling approach\cite{loeffler2020active,goldpaper,nielsen2013nested,varughese2024active} , we generate configurations spanning near-ground-state, metastable, and high-energy states, all re-labeled with DFT energies and forces. This comprehensive dataset challenges the EqNN to capture not only equilibrium energetics but also the structural distortions and defect-related physics characteristic of real materials. Such sampling strategies ensure that the resulting symbolic models are robust and transferable across diverse thermodynamic regimes.

We generate three distinct EAM potentials for aluminum using the EqNN framework to probe the impact of initialization and optimization pathways. The first model is trained from scratch using a combination of Monte Carlo Tree Search (MCTS) \cite{koneru2025development, iype2013parameterization,cosseddu2017force,patra2020accelerating,koneru2024machine, manna2022learning,
koneru2024ab} and gradient descent \cite{gilbert1992global}, enabling a global-to-local exploration of the functional landscape. The second and third models employ transfer learning from a pre-trained copper potential \cite{varughese2026physically}: one refined using MCTS plus gradient descent, and another using gradient descent alone. These transfer-learned variants exploit the structural similarity between copper and aluminum, reducing convergence time by over an order of magnitude compared to random initialization. Collectively, the three potentials yield distinct yet physically consistent symbolic expressions for the EAM components—each representing a different local optimum in the high-dimensional symbolic search space.

While all three models achieve comparable accuracy (sub-10 meV/atom energy error relative to DFT), they display complementary strengths across physical properties. The model with the random start reproduces phonon dispersions most faithfully, while the transfer learned models achieves superior surface energy predictions, and the elastic constants and strain-dependent energy curves. Recognizing that no single model uniformly dominates across all properties, we integrate the three potentials through ensemble symbolic regression—a weighted averaging of their closed-form equations. This ensemble approach yields a composite potential that leverages the strengths of each constituent model, significantly improving overall performance across multiple benchmarks. The ensemble potential exhibits enhanced agreement with DFT in equation-of-state curvature, phonon spectra, surface energies, and elastic moduli, while maintaining analytical interpretability and computational efficiency.

Relative to the state-of-the-art machine-learning potentials, the proposed ensemble symbolic framework achieves a unique balance of accuracy, transparency, and generalizability. Unlike purely neural architectures, our model reveals explicit mathematical dependencies between atomic and electronic variables, allowing users to inspect, modify, or extend the potential form as scientific understanding evolves. Compared to conventional symbolic regression, it leverages gradient-based learning and transferability to scale across elements and achieve rapid convergence. Most importantly, by incorporating far-from-ground state data and ensemble learning, it captures material behavior under realistic, non-equilibrium conditions—where many existing potentials, both empirical and ML-based, tend to break down. This combination of interpretability, robustness, and cross-regime fidelity represents a decisive step toward generalizable symbolic models that unify physics-based intuition with the flexibility of data-driven discovery.

Our work thus advances the state of the art in several key aspects. First, it introduces a physics-informed symbolic neural architecture capable of learning full EAM-like potentials that retain explicit analytic form. Second, it demonstrates the effectiveness of transfer learning and global-local optimization (MCTS + gradient descent) for accelerating symbolic model discovery. Third, it establishes the first ensemble symbolic regression framework for interatomic potentials, showing that combining diverse symbolic trajectories can systematically improve cross-property fidelity. Finally, it demonstrates that the resulting ensemble potential remains stable and predictive under far-from-equilibrium conditions, where conventional empirical and ML potentials often fail.

By uniting symbolic interpretability, data-driven adaptability, and non-equilibrium robustness, this hybrid symbolic–neural approach offers a new paradigm for developing interatomic potentials that are simultaneously accurate, transferable, and physically transparent. It bridges the gap between traditional analytical models and high-capacity ML architectures, demonstrating that closed-form equations derived from machine learning can serve as next-generation building blocks for interpretable, generalizable, and scalable materials simulations.

\section{METHODS}

\subsection{Neural Network Representation of the Embedded Atom Method}

We construct a machine-learning-based interatomic potential that strictly follows the Embedded Atom Method (EAM) formalism. The total energy is written as:
\begin{equation}
E_{\text{total}} = \sum_i \left[ \frac{1}{2} \sum_{j \neq i} \phi(r_{ij}) + F\left(\sum_{j \neq i} \rho(r_{ij})\right) \right]
\end{equation}\cite{sutton1990long}
where $\phi(r_{ij})$ is the pair potential, $\rho(r_{ij})$ is the density contribution from atom $j$ to atom $i$, and $F(\rho_i)$ is the embedding energy as a function of the accumulated local density $\rho_i=\sum_{j\neq i}\rho(r_{ij})$.

We represent the three EAM components using three EqNNs: one for the pair potential $\phi(r)$, one for the density contribution $\rho(r)$, and one for the embedding function $F(\rho)$. The pair and density EqNNs take scalar interatomic distances within the cutoff as inputs, while the embedding EqNN takes the accumulated scalar density $\rho_i$ as input.

Therefore the same learned one-dimensional pair and density functions are evaluated over all neighbor pairs and summed according to the EAM expression. This summation structure allows the potential to handle structures with different numbers of atoms and neighbors.

The three EqNNs are trained simultaneously within a single computational graph. Errors in the predicted total energy are backpropagated through the EAM expression to update the parameters of $\phi$, $\rho$, and $F$ concurrently. This preserves the physical EAM decomposition while allowing the functional forms of each component to be learned from data.

The density and embedding EqNNs contain one symbolic hidden layer between input and output. The pair EqNN contains two symbolic layers: the first layer generates inverse-power and exponential features of $r$, and the second layer applies either an identity or exponential activation to learned linear combinations of these features. This two-layer structure allows us to search for complex combinations of functions such as $\exp(a/r+b/r^6)$ or $A\exp(B/r^6+C/r+D\exp(Er))$ and they can emerge even though the primitive activation library contains only simpler functions.

For each atomic configuration, the interatomic distances between each atom and its neighbors within the cutoff radius are computed and passed to the pair and density EqNNs. The pair EqNN evaluates the pair contribution for each neighbor distance, while the density EqNN evaluates the corresponding density contribution. These density contributions are summed around each atom to form the local density input to the embedding EqNN. Finally, the pair-energy contributions and embedding-energy contributions are summed over all atoms according to the EAM expression to obtain the total energy of the structure.

The activation functions used in each EqNN are listed in Table~\ref{tab:basis_functions}.

\begin{table}[h!]
\centering
\caption{Library of basis functions used as activation functions in the three simultaneously trained EqNNs. The pair-potential EqNN contains two symbolic layers, while the density and embedding EqNNs contain one symbolic layer.}
\label{tab:basis_functions}
\begin{tabular}{|c|c|}
\hline
\textbf{EqNN Instance} & \textbf{Basis Functions} \\ \hline

\multirow{10}{*}{\textbf{Pair Potential ($\phi$)}} 
 & \textbf{Layer 1:} $\exp(-r)$ \\ \cline{2-2} 
 & $r^{-1.0}$ \\ \cline{2-2} 
 & $r^{-6.0}$ \\ \cline{2-2} 
 & $r^{-7.0}$ \\ \cline{2-2} 
 & $r^{-8.0}$ \\ \cline{2-2} 
 & $r^{-9.0}$ \\ \cline{2-2} 
 & $r^{-10.0}$ \\ \cline{2-2} 
 & $r^{-12.0}$ \\ \cline{2-2} 
 & $r^{-14.0}$ \\ \cline{2-2}
 & \textbf{Layer 2:} Identity, $\exp(\cdot)$ \\ \hline

\multirow{6}{*}{\textbf{Density ($\rho$)}} 
 & \textbf{Layer 1:} $\exp(-r)$ \\ \cline{2-2} 
 & $r^{-1.0}$ \\ \cline{2-2} 
 & $r^{-5.0}$ \\ \cline{2-2} 
 & $r^{-6.0}$ \\ \cline{2-2} 
 & $r^{-7.0}$ \\ \cline{2-2} 
 & $r^{-8.0}$ \\ \hline

\multirow{5}{*}{\textbf{Embedding Energy ($F$)}} 
 & \textbf{Layer 1:} $\rho^{0.5}$ \\ \cline{2-2} 
 & $\rho^{1}$ \\ \cline{2-2} 
 & $\rho^{2}$ \\ \cline{2-2} 
 & $\rho^{3.0}$ \\ \cline{2-2} 
 & $\rho^{4.0}$ \\ \hline

\end{tabular}
\end{table}

\subsection{Equation Learner Neural Network Architecture}

Equation learner neural networks (EqNNs) combine neural-network optimization with symbolic regression to learn explicit mathematical equations from data. In this work, EqNNs are used to obtain closed-form EAM functions that remain differentiable, computationally efficient, and directly inspectable.

Instead of conventional activations such as ReLU or sigmoid, each EqNN layer uses a predefined library of symbolic activations, including inverse powers, exponentials, square roots, and polynomial terms. The learnable coefficients multiply the outputs of these activations, so each symbolic layer represents a weighted combination of physically motivated basis functions.

For the density and embedding networks, this produces a single-layer symbolic expansion. For the pair network, the second symbolic layer acts on the output of the first layer, allowing function-of-a-function forms.

To train the EqNNs, we first use continuous-action Monte Carlo Tree Search (c-MCTS) to explore basis combinations and parameter values. Once a promising region is identified, gradient-based optimization is used to refine the active coefficients. This hybrid global--local strategy allows the search to explore a large symbolic space while still producing compact closed-form equations.

A key part of the EqNN procedure is the pruning of non-contributory basis functions. Because reciprocal and exponential functions can have large effects even when their coefficients are small, pruning based only on coefficient magnitude is unreliable. We therefore use a weight-toggling mechanism that maps a pseudo-weight $v_i$ to the real weight $w_i$:
\begin{equation}
    w_{i} = 
    \begin{cases}
        v_{i} + L_{\text{zone}}, & v_{i} \leq -L_{\text{zone}} \\
        0, & -L_{\text{zone}} \leq v_{i} \leq L_{\text{zone}} \\
        v_{i} - L_{\text{zone}}, & L_{\text{zone}} \leq v_{i}
    \end{cases}
\end{equation}
where $L_{\text{zone}}$ defines a user-controlled null zone around zero.

Larger values of $L_{\text{zone}}$ produce more aggressive pruning and favor sparser equations, whereas smaller values retain more terms for longer during the search. Thus, $L_{\text{zone}}$ is a hyperparameter that controls the sparsity--accuracy trade-off and should be selected based on the target system and the desired compactness of the final expression. We use a value of 0.125 as our value of $L_{\text{zone}}$.

Symbolic basis functions can also produce numerical instabilities, including NaN or infinite values, especially near singularities. To prevent optimizer failure and enforce physically admissible functions, we include penalty terms in the loss function for domain errors and for violation of the required positivity of the total evaluated pair and density functions within the cutoff range.

\subsection{Loss Function}
Development of an accurate symbolic model requires a comprehensive loss function that integrates constraints imposed by basis set functions and physical principles. Traditional loss functions often fail to account for domain-specific errors and physical constraints, leading to suboptimal model performance and potential numerical instabilities. Our modified loss function explicitly incorporates constraints arising from the basis set functions. These constraints are crucial for maintaining the physical realism of the model. In scenarios where the basis functions produce domain-related errors resulting in NaN or infinite values, the loss function assigns a significantly large penalty. This approach prevents the optimizer from crashing and ensures the stability of the training process. The EAM model necessitates that the pair and density functions yield positive outputs at all distances within the cutoff range. To enforce this, the loss function imposes a large penalty whenever any network outputs a negative value within the specified cutoff distance. 

The large penalty value is used as a hard rejection criterion during the symbolic search rather than as a smoothly balanced optimization weight. When a candidate expression violates the positivity or domain constraints, the penalty makes that branch of the MCTS search tree unfavorable, effectively pruning it and redirecting the search toward physically admissible regions of the expression space.

This strategy allows the EqNN to retain both positive and negative coefficients in individual symbolic terms while enforcing positivity only on the final evaluated pair and density functions. Therefore, individual components may have negative coefficients or negative contributions, provided that the complete learned function remains positive over the sampled distance range up to the cutoff.
This constraint is vital for maintaining the physical integrity of the model, as negative values could lead to non-physical predictions. Empirical observations and theoretical insights indicate that structures near their equilibrium configurations are more prevalent within the temperature ranges of interest. These structures also play a critical role in determining properties such as elastic constants. Therefore, our loss function assigns higher weights to these near-equilibrium structures. This weighting scheme enhances the model's predictive efficiency, particularly for properties that are sensitive to structural variations. To quantify the accuracy of energy predictions, we use the mean absolute error (MAE) as our primary metric. Our comprehensive loss function is formulated as follows:


\begin{eqnarray}
\text{Loss Function} &=& w \cdot \text{MAE}_{\text{NE}} 
+ \text{MAE}_{\text{RE}} \nonumber \\
&&+ \text{Penalty}_{\text{Pair}} 
+ \text{Penalty}_{\text{Density}} \nonumber \\
&&+ \text{Penalty}_{\text{Domain}}~.
\end{eqnarray}
\[
  Penalty_{Pair}= 
\begin{cases}
  1e18 & Pair(r) < 0 \\
  0,      & Pair(r) > 0 
\end{cases}
\]
\[
  Penalty_{Density}= 
\begin{cases}
  1e18 & Density(r) < 0 \\
  0,      & Density(r) > 0 
\end{cases}
\]
\[
  Penalty_{Domain}= 
\begin{cases}
  1e18 & NaN/Inf \ outputs \\
  0,      & otherwise
\end{cases}
\]

where:
\begin{itemize}
  \item $w$ is the weighting factor for near-equilibrium structures,
  \item $\text{MAE}_{\text{NE}}$ is the mean absolute error for near-equilibrium structures,
  \item $\text{MAE}_{\text{RE}}$ is the mean absolute error for the rest of the structures,
  \item $\text{Penalty}_{\text{Pair}}$ is the penalty for the pair function producing negative values within the cutoff,
  \item $\text{Penalty}_{\text{Density}}$ is the penalty for the density function producing negative values within the cutoff,
  \item $\text{Penalty}_{\text{Domain}}$ is the penalty for domain errors in the basis set functions.
\end{itemize}

\subsection{Nature of the Data Set - Training, Validation and Test Data Set Generation}

We employed two sets of datasets – (i) we generated training and test dataset using an active learning approach to sample configurations for benchmarking the EqNN by recovering the known functional forms (SC-EAM) and (ii) the energies of the sampled configurations in (i) were recalibrated with DFT calculations and used to generate new symbolic regression (SR) models that work well across a broad range of energies, from near to far-from-ground state. The details for the DFT calculations are provided in supplementary table S1. 

In our active learning workflow, the sampling of the configurations is performed using the nested ensemble sampling method. This approach efficiently samples structural configurations across the different energy ranges, providing a robust dataset for model training. We begin by generating a large number of initial configurations of the system, ensuring that these configurations uniformly covered the entire phase space. In a system consisting of N particles, random positions were generated for each particle within the defined boundaries. During the initial sampling procedure, all generated configurations were accepted regardless of their energies. The potential energy E for each configuration was computed using the relevant potential energy function, ie the current state of the symbolic NN that is being trained.The energies of all accepted configurations were recorded, and the median energy E$_1$ was calculated from these values. This median energy served as the threshold for the next iteration. The configuration and corresponding energy information were stored in the structure set, ensuring a comprehensive record of the initial sampling phase.

For subsequent iteration, the median energy from the first iteration E$_1$ was used as the energy threshold. New configurations were generated by perturbing the accepted configurations from the previous iteration, and only those with energies lower than E$_1$ were accepted. The energies of these configurations were recorded, and a new median energy E$_2$ was computed. This iterative process continued, with each iteration using the median energy from the previous iteration as the new energy threshold. For each iteration n the median energy E$_{n-1}$ was recorded along with the configuration and energy values. The final structure set was compiled by combining the configuration and energy data from all iterations. This compilation ensured that the structure set included a diverse range of configurations with corresponding energy values covering the relevant phase space comprehensively. 

We utilize this final training set for discovering new models for Aluminum by recalculating their energy values using Density Functional Theory (DFT)\cite{hohenberg1964inhomogeneous} using the Vienna Ab initio Simulation Package (VASP)\cite{kresse1996efficiency} and using that recalculated data for the training data for the equation search. 

\subsection{EqNN Training- Global Search and Transfer Learning}

\begin{figure*}[ht]
  \centering
  \includegraphics[width=0.8\textwidth]{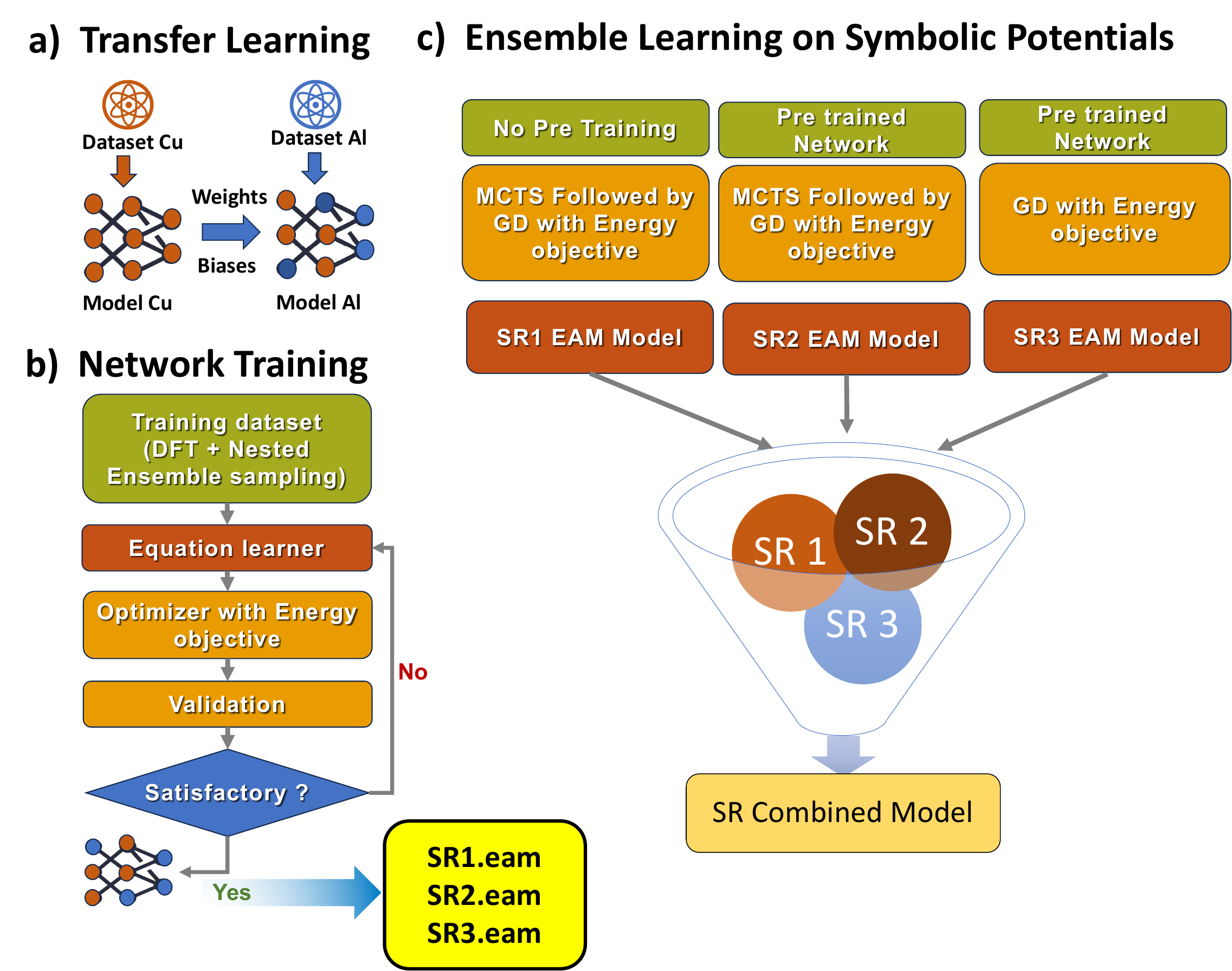}
  \caption{
  \textbf{Schematic illustration of the multi-stage symbolic regression framework for ensemble learning of interatomic potentials.}
  (a) \textbf{Transfer learning:} Pre-trained parameters (weights and biases) from a Cu model are transferred to initialize an Al model, enabling efficient reuse of learned atomic interaction representations. 
  (b) \textbf{Network training:} A closed-loop workflow combining nested sampling with an equation learner, energy-based optimization, and validation ensures physically consistent potential generation. The iterative loop continues until convergence criteria are satisfied. 
  (c) \textbf{Ensemble learning on symbolic potentials:} Three distinct training strategies—no pre-training (SR1 ), pre-training with Monte Carlo Tree Search (MCTS) and gradient descent (SR2), and pre-training with gradient descent only (SR3)—produce three different complementary symbolic potentials. These are aggregated to form an ensemble model (SR\_Combined), which integrates their strengths to achieve improved transferability and near-DFT fidelity across diverse configurations.
  }
  \label{fig:ensemble_sr_framework}
\end{figure*}

The optimization of Equation Neural Networks (EqNNs) as described in Fig \ref{fig:ensemble_sr_framework} was carried out using a hybrid global--local strategy designed to balance exploratory search with fine-grained refinement of model parameters. The global phase employed a continuous-action variant of Monte Carlo Tree Search (c-MCTS \cite{manna2022learning,koneru2023multi}), which extends the classical discrete MCTS formalism to operate in continuous, high-dimensional parameter spaces. In this framework, each node represents a hypersphere in parameter space and the exploration--exploitation balance is maintained by dynamically adjusting the search radius as the depth of the tree increases. To address degeneracy and inefficiency inherent to large, nonconvex landscapes, c-MCTS incorporates three key mechanisms: (i) a uniqueness function to suppress redundant sampling, (ii) a depth-correlated geometric scaling of the search radius with a rate of 0.5 and a depth limit of 25, and (iii) adaptive playout sampling (playout size of 10) that biases sampling toward promising regions. All parameters in the global phase were constrained within a bounded domain of $\pm 5.0$ to ensure both sufficient exploration and numerical stability. The Hyperparameters used for MCTS are provided in Supplementary Table S2

Two global-search initialization pathways were explored. The first used random sampling to initialize the search, forming the basis of what we designate as the SR1 workflow. The second employed a previously trained copper EqNN model \cite{varughese2026physically} as the starting point for c-MCTS, constituting the first form of transfer learning used in this work. This copper-based initialization—forming the SR2 workflow—substantially narrowed the effective region of interest in the early stages, enabling faster descent toward low-loss basins compared to the random-start SR1 approach. In both cases, the global search efficiently reduced the loss to within a few hundred meV\,atom$^{-1}$, but convergence slowed as the optimization landscape tightened, highlighting the diminishing returns of random playout sampling in increasingly narrow hyperspheres.

The workflow transitioned to local optimization once c-MCTS reached a loss of approximately $150$~meV\,atom$^{-1}$ or stagnated for 5{,}000 steps. In this phase, gradient-based refinement was performed using the Adam optimizer with a learning rate of $10^{-4}$. Only active (non-masked) weights were updated, allowing the optimizer to fine-tune the real-valued parameters underlying the EqNN’s symbolic structure. This refinement consistently improved the accuracy beyond what global sampling alone could achieve.

A second, more direct form of transfer learning was also implemented. Here, the copper EqNN was not used to initialize the c-MCTS phase; instead, it was directly fine-tuned using gradient descent from the first step. This workflow—designated as SR3—provided the fastest convergence among all three strategies (SR1: random$\rightarrow$c-MCTS$\rightarrow$Adam; SR2: Cu$\rightarrow$c-MCTS$\rightarrow$Adam; SR3: Cu$\rightarrow$GD). The strong inductive bias encoded in the copper EqNN enabled Adam, operating at the same learning rate of $10^{-4}$, to rapidly descend toward high-quality minima without the need for a global exploratory stage.

Taken together, this hybrid methodology demonstrates that combining the exploratory power of c-MCTS with the fine-resolution capability of gradient-based optimization yields a robust pipeline for discovering accurate and interpretable symbolic expressions. Moreover, both forms of transfer learning—initialization via the Cu EqNN in the global phase and direct Adam-based fine-tuning of the Cu model—significantly accelerate convergence and highlight the adaptability of EqNNs when prior knowledge is available.

\subsection{Greedy Model Pruning}
Once the global and local search strategies reduce the loss below 10~meV~atom$^{-1}$, we evaluate the resulting symbolic models. Although the $L_{\text{zone}}$--based pruning scheme removes activation functions that contribute little to loss reduction, the resulting expressions often remain algebraically bulky, with the potential energy surface represented as a distributed superposition of many weakly contributing terms. This occurs because the optimization process is still driven by entropy in function space: many distinct combinations of basis functions can achieve nearly identical loss values, leading to unnecessarily complex representations. To overcome this limitation and recover minimal, physically interpretable forms, we therefore apply a subsequent greedy model-order reduction procedure that explicitly targets algebraic simplicity while preserving quantitative accuracy.

 To obtain compact, interpretable analytic forms, we employed a greedy model-order reduction strategy applied directly to the symbolic EAM equations. Starting from the full fitted expression, each candidate basis term was removed one at a time and the resulting truncated model was evaluated against the original reference equation over the physically relevant interatomic-distance range $r \in [1,7]$~\AA. The term whose removal produced the smallest increase in error was provisionally discarded, after which the remaining coefficients were refitted by nonlinear least squares to minimize the residual. This prune--refit cycle was repeated iteratively, progressively eliminating redundant terms while preserving quantitative fidelity. The procedure was terminated when either removing any additional term caused the error with respect to the original function to exceed $10^{-1}$, or when refitting a further reduced model could no longer achieve an error with respect to the original function below $10^{-3}$. In this way, we systematically identified the minimal functional form that reproduces the original equation to high accuracy over the data-supported domain, yielding compact, physically interpretable expressions without loss of predictive quality.

\subsection{Melting point from two--phase coexistence}

The melting temperature ($T_\mathrm{m}$) of Aluminum was determined using a high-fidelity two-phase (solid–liquid) coexistence protocol implemented in LAMMPS\cite{LAMMPS}, applied to simulation cells containing approximately $3\times10^{5}$ atoms and evolved for \SI{1}{\nano\second}. A relaxed FCC supercell was first expanded to construct a slab geometry sufficiently large to host two solid–liquid interfaces under periodic boundary conditions. The system was partitioned into crystalline and molten halves; the liquid region was generated by heating to \SI{2500}{\kelvin} while the solid region was held fixed by zeroing atomic forces to preserve crystallinity. Following complete melting, the liquid portion was quenched to a trial temperature $T_{\mathrm{trial}}$ near the expected $T_\mathrm{m}$ to minimize density discontinuities, after which the composite slab was equilibrated to remove residual stresses. Coexistence simulations were then performed in the isothermal–isobaric ensemble at 1 bar using a Nosé–Hoover thermostat–barostat with a \SI{2}{\femto\second} timestep. Trajectories were obtained over a temperature grid from \SI{700}{\kelvin} to \SI{1200}{\kelvin}. At the end of each simulation, the structural identity of every atom was determined using common-neighbour analysis, enabling us to quantify the final percentage of FCC and liquid atoms.  At any fixed temperature, the thermodynamically favoured phase naturally grows at the expense of the other, allowing the melting point to be identified from the temperature at which neither phase exhibits net growth. This full workflow was repeated independently for the SR1, SR2, SR3, and SR\_Combined symbolic regression potentials.

\section{Results}

To evaluate the capabilities of symbolic learning for interatomic potentials, we examine how the Equation Neural Network (EqNN) framework recovers and extends the analytical structure of the Embedded Atom Method (EAM) for aluminum. Starting from a validation exercise against the canonical Sutton–Chen potential, we show that the EqNN systematically rediscovers functional forms that reproduce reference energies and forces with sub - 10 meV/atom errors.After demonstrating that the EqNN can exactly rediscover the analytical structure of the Sutton–Chen EAM, we next use the framework to learn the DFT potential energy landscape of Aluminum. To explore how initialization influences the symbolic search trajectory, we train three EqNN models under distinct starting conditions: (i) fully random initialization, (ii) transfer initialization guided by Monte Carlo Tree Search (MCTS), and (iii) transfer initialization guided by gradient descent. When each model is optimized on the DFT dataset, they converge to different—but equally physically meaningful—analytic forms. Although all three models are optimized on the same DFT data, the resulting symbolic potentials ultimately exhibit different strengths—one reproduces phonon dispersions most accurately, another yields the best surface energetics, and the third provides the most accurate elastic constants. These complementary characteristics highlight the diversity of solutions within the symbolic search space and motivate the construction of an ensemble symbolic potential that integrates the three EqNNs to achieve superior agreement with DFT benchmarks. The ensuing results section presents detailed quantitative comparisons, showing that our ensemble model delivers enhanced accuracy and robustness across near-ground state and far-from-ground state energy regimes, including equation-of-state behavior, phonon spectra, surface energies, and elastic constants, while maintaining full analytical interpretability.

\subsection{Benchmarking - Recovery of Sutton Chen EAM}

We benchmarked and validated our symbolic regression framework by reconstructing the potential energy surface (PES) of Aluminum as defined by the Sutton–Chen embedded atom model (SC-EAM). The structural dataset was initialized with bulk equation-of-state configurations and low-index surface structures, which were subsequently expanded through an active learning workflow employing nested ensemble sampling. At each sampling iteration, a structural similarity analysis based on the Smooth Overlap of Atomic Positions (SOAP) descriptor followed by principal component analysis (PCA)\cite{abdi2010principal} was performed to identify configurations that were both energetically and structurally distinct from the existing dataset. These dissimilar structures were selectively incorporated into the training pool to maximize configurational diversity and ensure uniform coverage of the potential energy landscape. The reference energies and atomic forces for all configurations were obtained from the SC-EAM potential, enabling the Equation Neural Network (EqNN) to progressively learn and reproduce the underlying analytic form of the Sutton–Chen potential with high fidelity.

\begin{figure*}[ht!]
  \centering
  \includegraphics[width=0.95\textwidth]{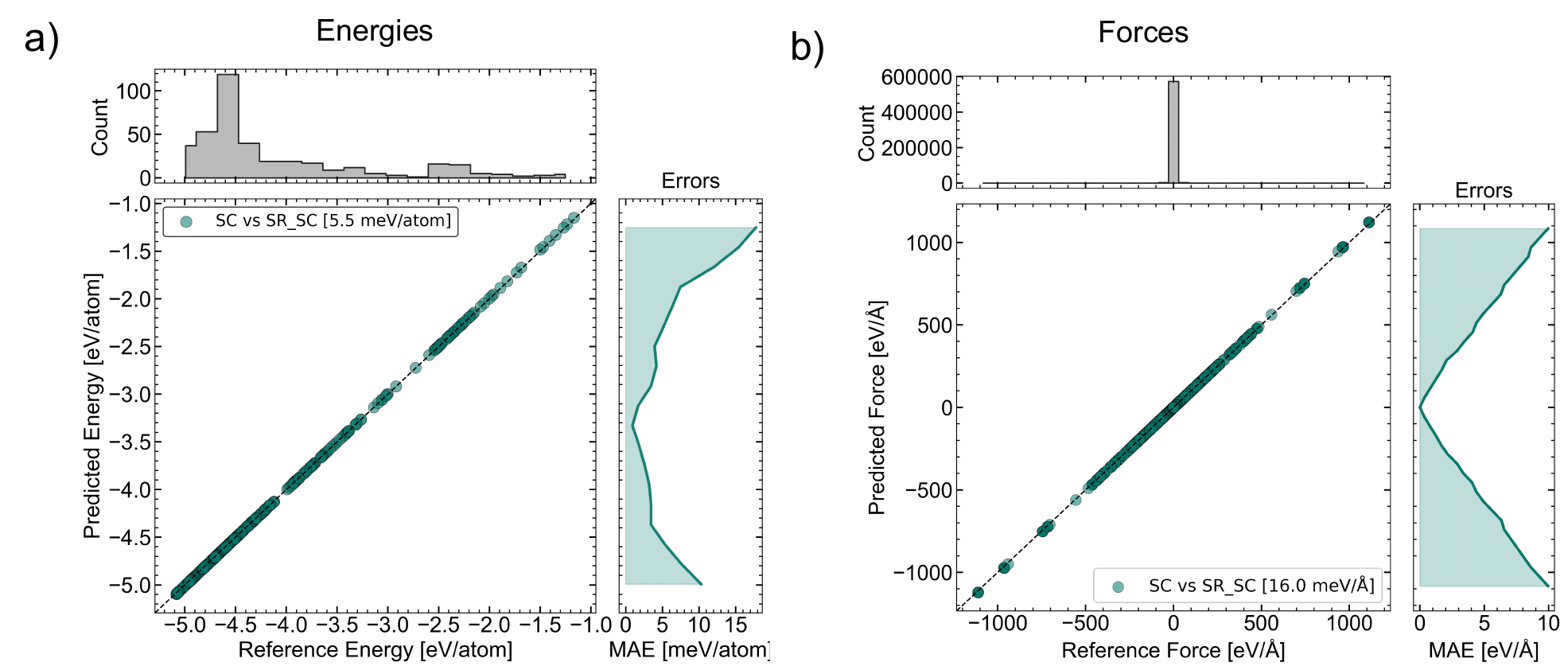}
  \caption{\textbf{Correlation between SC-EAM reference and symbolic regression (SR) predicted energies and forces for Aluminum.} 
  The Equation Neural Network (EqNN) model was trained to reproduce the Sutton–Chen Embedded Atom Model (SC-EAM) functional form for Aluminum using a progressively enriched dataset comprising bulk and surface configurations. The dataset was expanded through nested ensemble sampling, with SOAP–PCA analysis employed at each iteration to identify structurally and energetically diverse configurations. 
  The SR model yields mean absolute errors (MAEs) of 8.9 and 8.1~meV/atom for the training and test energies, respectively, and 8.4 and 7.9~meV/\AA{} for the corresponding forces. 
The tight clustering of data points along the identity line (dashed diagonal) demonstrates that the symbolic model accurately recovers the reference potential energy surface, effectively mapping both the scalar energy landscape and the force gradients.}
  \label{fig:Al_EnergyForceCorr}
\end{figure*}

To assess the fidelity of symbolic reconstruction, we directly compared the recovered analytic expressions with the reference SC-EAM formulation. The original Sutton–Chen expressions for Aluminum are:
\begin{eqnarray}
E_{\text{pair}}(r_{ij}) &=& \frac{296.209798418066}{r_{ij}^{7}}, \nonumber \\
\rho(r_{ij}) &=& \frac{1303.92697581447}{r_{ij}^{6}}, \nonumber \\
F_{\text{emb}}(\rho) &=& -1.0\,\sqrt{\rho},
\end{eqnarray}

while the symbolic expressions discovered by the EqNN are:

\begin{eqnarray}
E_{\text{pair}}(r_{ij}) &=& \frac{298.553388211222}{r_{ij}^{7}}, \nonumber \\
\rho(r_{ij}) &=& \frac{1285.64001212281}{r_{ij}^{6}}, \nonumber \\
F_{\text{emb}}(\rho) &=& -1.016577172082\,\sqrt{\rho}.
\end{eqnarray}

The EqNN identified the analytic exponents of the original SC-EAM formulation, indicating that the model captures the underlying functional dependence of the potential in addition to its numerical accuracy. The prefactor magnitudes deviate by less than 2\% from the reference model, while the embedding coefficient differs by only $\sim$1.6\%, preserving the characteristic square-root form of the embedding energy. Quantitatively, the SR model achieves mean absolute errors (MAEs) of 8.9 and 8.1~meV/atom for the training and test energies, and 8.4 and 7.9~meV/Å for the corresponding forces, confirming that both the energetic and force landscapes of the Sutton–Chen potential are faithfully reproduced. As shown in Fig.~\ref{fig:Al_EnergyForceCorr}, the SR-predicted values exhibit near-perfect correlation with the SC-EAM references, validating that the EqNN-based symbolic regression framework can autonomously discover compact, interpretable, and physically accurate analytic potentials directly from data. (The performance comparison for SC EAM with respect to DFT is provided in Supplementary Figures S1 and S2).
The force comparison in Fig.~2 corresponds to the Sutton--Chen EAM potential and its EqNN-equivalent representation, not to the final DFT-trained symbolic-regression potentials. The large predicted forces arise when the Sutton--Chen form, which was primarily parameterized for ground-state and near-ground-state properties, is evaluated on highly distorted configurations outside its original fitting regime.

Thus, the broad force range in Fig.~2 reflects the limited transferability of the reference Sutton--Chen potential to far-from-ground state structures, rather than an instability introduced by the final symbolic-regression models.

\subsection{Transfer Learning and Symbolic Potential Discovery for Aluminum}

All generated configurations were subsequently evaluated using density functional theory (DFT) to obtain reference energies and atomic forces for the next stage of training. To construct the Equation Neural Network (EqNN) potential for Aluminum, we explored three distinct initialization and optimization strategies designed to evaluate the impact of prior knowledge transfer on the convergence and accuracy of symbolic regression.

First, we trained an EqNN model initialized with random weights, following the same Monte Carlo Tree Search (MCTS) and gradient-based optimization protocol as employed in the copper study. This served as the baseline, denoted as SR1. Second, we leveraged the best-performing copper EqNN model and refined it for Aluminum using both MCTS and gradient descent, a transfer learning strategy incorporating symbolic exploration (SR2). Finally, we initialized the Aluminum EqNN with the same pre-trained copper model but performed only gradient-based optimization, thereby testing transfer learning without global search (SR3).

\begin{figure*}[ht!]
  \centering
  \includegraphics[width=0.85\linewidth]{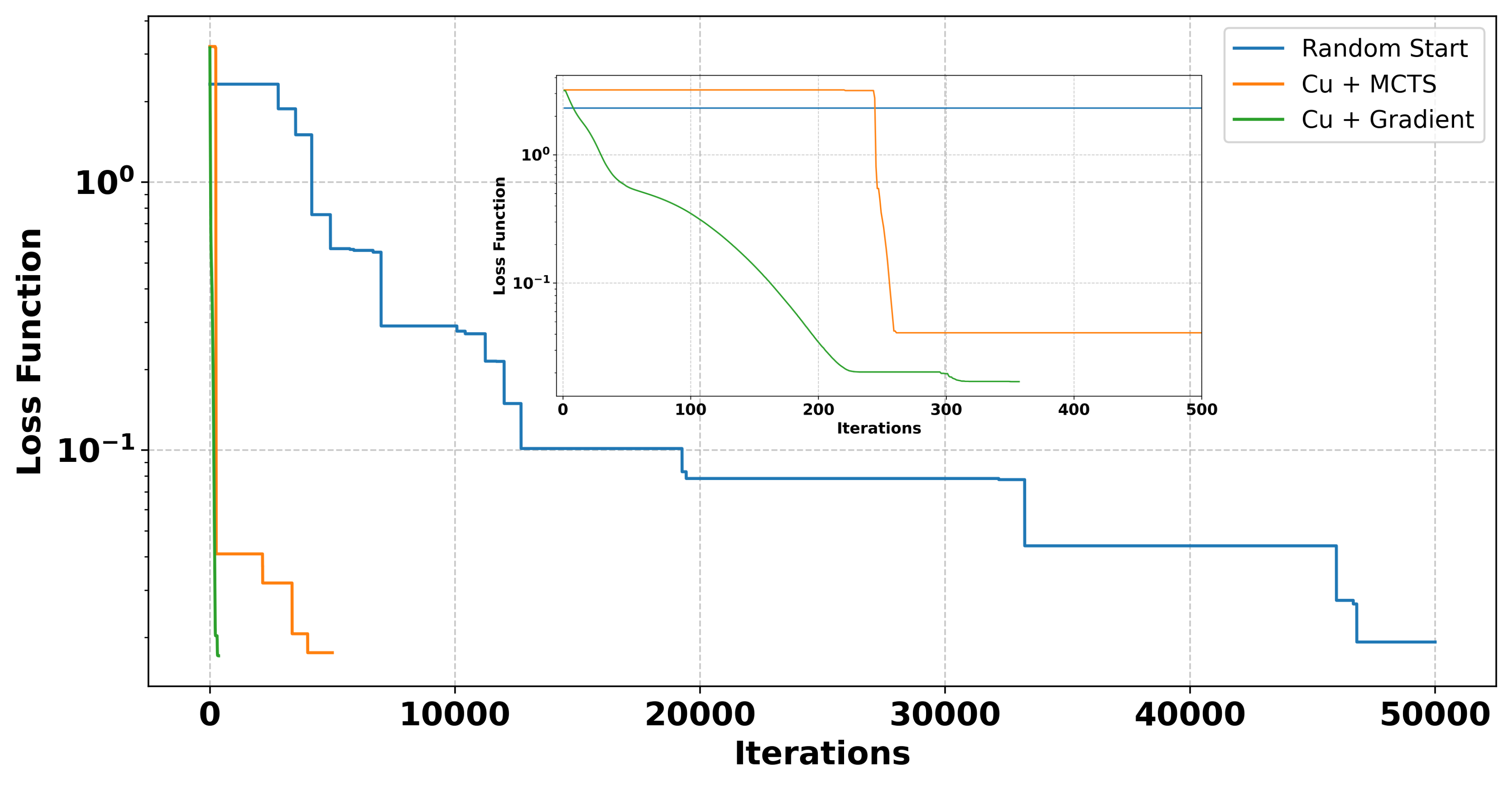}
  \caption{\textbf{Convergence behavior of EqNN training for Aluminum under three initialization strategies.} 
  Shown are validation losses (in meV/atom) as a function of optimization steps for randomly initialized training (blue), transfer learning from copper followed by MCTS + gradient descent (orange), and transfer learning from copper with gradient descent only (green). 
  The randomly initialized model required over 50{,}000 MCTS steps to converge, whereas transfer learning via MCTS + gradient descent converged within $\sim$5{,}000 steps. The model refined solely by gradient descent achieved convergence in fewer than 500 steps, underscoring the efficiency of symbolic transfer learning in accelerating potential discovery.}
  \label{fig:al_transfer_convergence}
\end{figure*}

Figure~\ref{fig:al_transfer_convergence} illustrates the convergence behavior of the three strategies. The randomly initialized EqNN exhibited the slowest convergence, requiring over 50{,}000 optimization steps to reach stability. In contrast, initializing from the pre-trained copper model and applying MCTS + gradient descent reduced the training time by nearly an order of magnitude. Transfer learning with gradient optimization achieved convergence in fewer than 500 steps. This efficiency underscores the reusability of symbolic representations derived from chemically similar systems and indicates the transferability of EqNN-based symbolic frameworks. All the three models were pruned further using the greedy model pruning protocol specified in the methods to simplify the model and enhance the intrerpritability of the model without sacrificing model performance.

Each of the three training strategies yielded distinct symbolic equations for the pair potential, electron density, and embedding energy. While all retained the embedded-atom form, the recovered coefficients and polynomial terms differed, reflecting the influence of initialization and search trajectory on the symbolic discovery process:

\paragraph{EqNN trained from random initialization (SR1):}

\begin{align}
E_{\text{pair}}(r_{ij}) &= 
0.450487 \exp\Bigg(
\frac{0.408279}{r_{ij}^{6}}
+ \frac{5.924174}{r_{ij}} \notag \\
&\qquad
- 0.068368 \exp(1.504135\, r_{ij})
\Bigg), \notag \\
\rho(r_{ij}) &=
\frac{23.453896}{r_{ij}^{6}}
+ \frac{1.091162}{r_{ij}^{5}} \notag \\
&\qquad
+ 6.453485 \exp(-0.758860\, r_{ij}), \notag \\
F_{\text{emb}}(\rho) &=
-1.6323 \sqrt{\rho}
- 0.00207 \rho^2
+ 0.2114 \rho .
\end{align}

\paragraph{EqNN initialized from copper model and optimized using MCTS + gradient descent (SR2):}

\begin{align}
E_{\text{pair}}(r_{ij}) &= 
0.085776 \exp\Bigg(
-\frac{0.686828}{r_{ij}^{6}}
+ \frac{8.749128}{r_{ij}} \notag \\
&\qquad
- 0.029743 \exp(1.658202\, r_{ij})
\Bigg), \notag \\
\rho(r_{ij}) &=
\frac{19.524875}{r_{ij}^{7}}
+ \frac{213.889468}{r_{ij}^{6}} \notag \\
&\qquad
+ 2.900293 \exp(-1.087303\, r_{ij}), \notag \\
F_{\text{emb}}(\rho) &=
-2.0594 \sqrt{\rho}
- 0.00463 \rho^2
+ 0.2708 \rho .
\end{align}

\paragraph{EqNN initialized from copper model and optimized using gradient descent only (SR3):}

\begin{align}
E_{\text{pair}}(r_{ij}) &= 
0.043653 \exp\Bigg(
-\frac{4.961811}{r_{ij}^{6}}
+ \frac{10.301217}{r_{ij}} \notag \\
&\qquad
- 0.034701 \exp(1.540214\, r_{ij})
\Bigg), \notag \\
\rho(r_{ij}) &=
\frac{13.687408}{r_{ij}^{7}}
+ \frac{195.064041}{r_{ij}^{6}} \notag \\
&\qquad
+ 2.935966 \exp(-1.169265\, r_{ij}), \notag \\
F_{\text{emb}}(\rho) &=
-2.1122 \sqrt{\rho}
- 0.00503 \rho^2
+ 0.2375 \rho .
\end{align}

\begin{figure*}[ht!]
  \centering
  \includegraphics[width=0.95\linewidth]{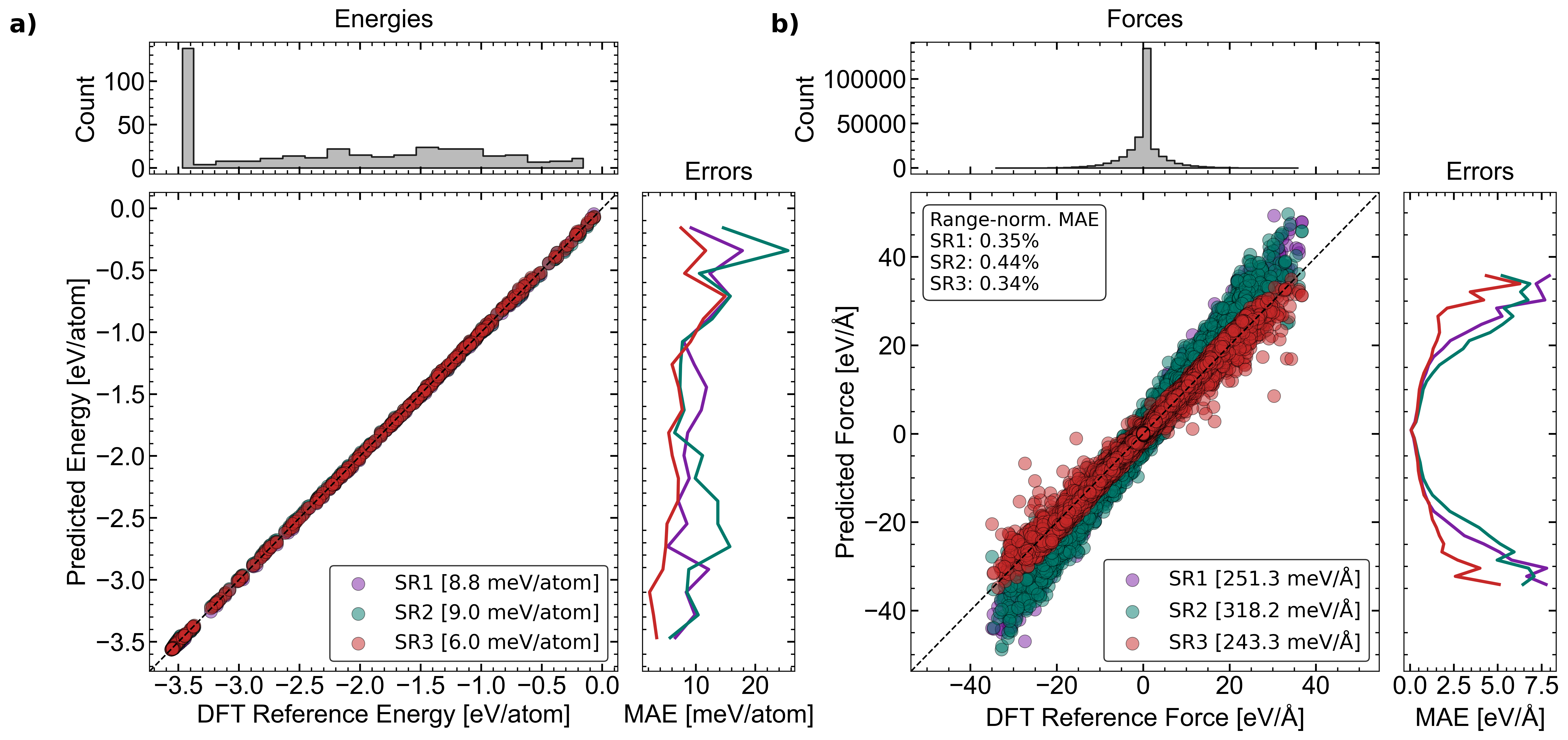}
  \caption{\textbf{Comparison of EqNN-predicted energies and forces for Aluminum against DFT references.} 
  (a) Energy and (b) force correlations for the three EqNN models—SR1 (random initialization), SR2 (Cu + MCTS), and SR3 (Cu + Gradient). 
  The upper panels show the DFT reference data distributions, while the right panels display the mean absolute error (MAE) distributions across the energy and force spectra. 
  All three models show strong parity with DFT predictions, with mean absolute errors of 8.9, 9.0, and 6.0~meV/atom for energy and 251.3, 318.2, and 243.3~meV/Å for forces, respectively. 
  The histograms reveal a pronounced peak near the ground state due to the inclusion of equilibrium, EOS, and surface structures, followed by an approximately uniform spread across higher-energy configurations. 
  The error profiles highlight distinct error localization behaviors across the three models. (For clarity, the separated energy correlation plots and complementary performance metrics are presented in Supplementary Figures S3–S5)}
  \label{fig:al_dft_corr}
\end{figure*}

Figure~\ref{fig:al_dft_corr} presents the energy and force correlation plots comparing EqNN predictions with DFT reference data for the three Aluminum models. All models exhibit excellent agreement with DFT, as reflected by the tight clustering of points along the parity line and the narrow error distributions. Notably, the SR3 model, obtained purely through gradient-based refinement from the pre-trained copper EqNN, achieved the lowest errors (6.0~meV/atom for energy and 243.3~meV/Å for forces), demonstrating that transfer learning from chemically analogous systems can yield both higher accuracy and faster convergence.

Figure~\ref{fig:al_dft_corr} top panels show the distribution of the structures used for training these networks. All datasets exhibit a pronounced concentration of structures near the ground state, as reflected in the histogram peak at low energies, corresponding to equilibrium, equation-of-state, and surface configurations. Beyond this region, the data are nearly uniformly distributed, ensuring balanced sampling across diverse configurations.

The error profiles (right panels) provide a detailed view of how predictive deviations evolve across the configurational landscape. The SR3 model shows the lowest and most stable errors, maintaining deviations below 4–5~meV/atom in the vicinity of the ground state and gradually increasing for structures further away from equilibrium. In contrast, SR2 and SR1 display non-monotonic error distributions, with error maxima localized near the ground state and at the farthest high-energy configurations, while exhibiting a relative dip in the intermediate energy range. This pattern suggests that while all three symbolic models capture the overall energy–force landscape with high fidelity, their predictive precision varies systematically across different regions of the potential energy surface, reflecting the influence of initialization and optimization trajectory on the learned symbolic forms.

Together, these results highlight the power of symbolic regression in learning physically interpretable interatomic potentials while maintaining quantitative agreement with quantum-mechanical references. The convergence acceleration and performance gains achieved through symbolic transfer learning demonstrate a scalable route toward cross-element generalization of interpretable potential models.

\subsection{Bulk and Surface Property Evaluation of Symbolic Aluminum Potentials}

\begin{figure*}
  \centering
  \includegraphics[width=0.85\linewidth]{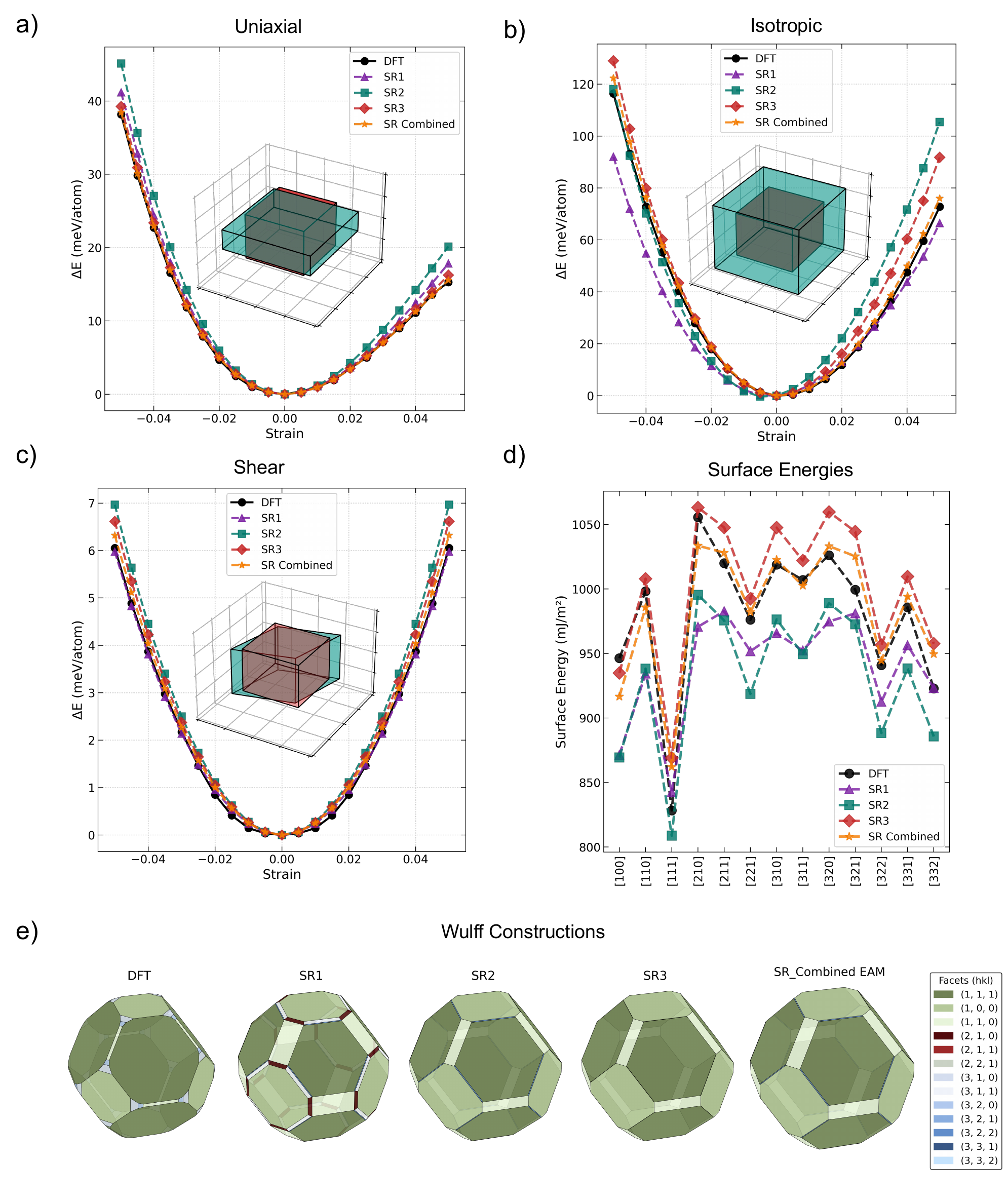}
  \caption{\textbf{Benchmarking of symbolic Aluminum potentials against DFT reference data.}
  Top and middle panels: equation-of-state (EOS) curves for fcc Aluminum under three distinct deformation modes—(i) uniaxial strain along one lattice axis at constant volume, (ii) isotropic volumetric compression and expansion, and (iii) simple shear deformation. Insets illustrate representative deformation geometries. 
  Bottom-right panel: surface energy comparison for low- and high-index facets. 
  Bottom row: corresponding Wulff constructions derived from the computed surface energies, illustrating the predicted equilibrium morphologies. 
  DFT reference data are shown in black, with SR1 (purple), SR2 (green), SR3 (red), SR Combined EAM (yellow).}
  \label{fig:Al_Bulk1}
\end{figure*}

To further validate the symbolic EqNN-derived Aluminum potentials, we examined their response under three deformation modes—uniaxial, isotropic volumetric , and shear—together with surface energetics and Wulff constructions, as summarized in Fig.~\ref{fig:Al_Bulk1}. Each test quantifies how well the symbolic potentials reproduce the DFT reference energy landscape under progressively non-equilibrium structural perturbations. Although all three symbolic regression–derived EqNN models (SR1, SR2, and SR3) achieve excellent energy prediction performance—each with a mean absolute error of less than 10~meV/atom—their ability to replicate other key physical properties varies significantly.

Under uniaxial deformation (top-left panel), all three symbolic potentials reproduce the characteristic parabolic energy–strain relation observed in DFT. 
Among them, SR3 (red) exhibits the best overall quantitative agreement, tracking the DFT curve across both tension and compression with minimal deviation. 
SR1 (purple) remains close to DFT up to strains of about $\pm0.03$ before showing slight over-stiffening, whereas SR2 (green) deviates most noticeably beyond $\sim$3\% strain, underpredicting the energy at large distortions. 
These differences indicate that SR3 yields the most balanced elastic and anharmonic response under anisotropic loading.

For the isotropic volumetric strain EOS (top-right panel), all models correctly reproduce the equilibrium volume and cohesive energy of fcc Aluminum. 
At large compressions, SR2 aligns most closely with DFT, whereas in the range from 0.025 compression to 0.015 expansion, SR3 provides the best match. 
Beyond $\sim$1.5\% tensile strain, SR1 becomes the most accurate, maintaining the correct curvature across the extended volume range. 
Together, these observations highlight that each symbolic model reproduces the bulk equation of state accurately in different strain regimes, reflecting subtle trade-offs between pair and embedding stiffness terms.

In the shear deformation case (middle-left panel), all three models capture the correct parabolic dependence on strain. 
SR1 most closely follows the DFT energy variation across the entire shear range, while SR2 and SR3 predict slightly stiffer responses at higher shear strains. 
At small strains ($|\varepsilon|<0.015$), all symbolic models produce marginally higher curvature than DFT, indicating a mild overprediction of the elastic constant $C_{44}$.

The surface energy comparison (bottom-right panel) shows that both SR1 and SR2 systematically underpredict surface energies across most orientations, although they correctly reproduce the energetic ordering of the low-index facets ($\gamma_{(111)} < \gamma_{(100)} < \gamma_{(110)}$). In contrast, SR3 provides the closest overall agreement with DFT, with modest overestimation for some higher-index surfaces and good correspondence for the low-index facets. The surface structures used in this analysis were taken from the Crystalium database~\cite{tran2016surface}, ensuring consistency with standardized first-principles surface calculations. The deviations observed for all models remain modest and within the expected range for analytic EAM-type potentials, indicating that the symbolic functional forms retain reasonable quantitative transferability from bulk properties to surface energetics.

The resulting Wulff constructions (bottom row) illustrate the equilibrium morphologies predicted from the respective surface energy anisotropies. 
All three models produce the characteristic truncated octahedral geometry dominated by \{111\} and \{100\} facets, consistent with the experimentally observed morphology of fcc Aluminum. 
SR3’s smoother anisotropy results in a more compact shape, whereas SR1 and SR2 display slightly enlarged \{100\} faces, consistent with their lower predicted $\gamma_{(100)}$ values. 
Overall, the close correspondence between the symbolic potentials and DFT across uniaxial, volumetric, and shear deformations—as well as surface energetics—confirms that the learned analytic forms accurately capture both the elastic and surface thermodynamic behavior of Aluminum.
\subsection{Phonon Dispersion and Lattice Dynamics}

\begin{figure*}[ht!]
  \centering
  \includegraphics[width=0.85\linewidth]{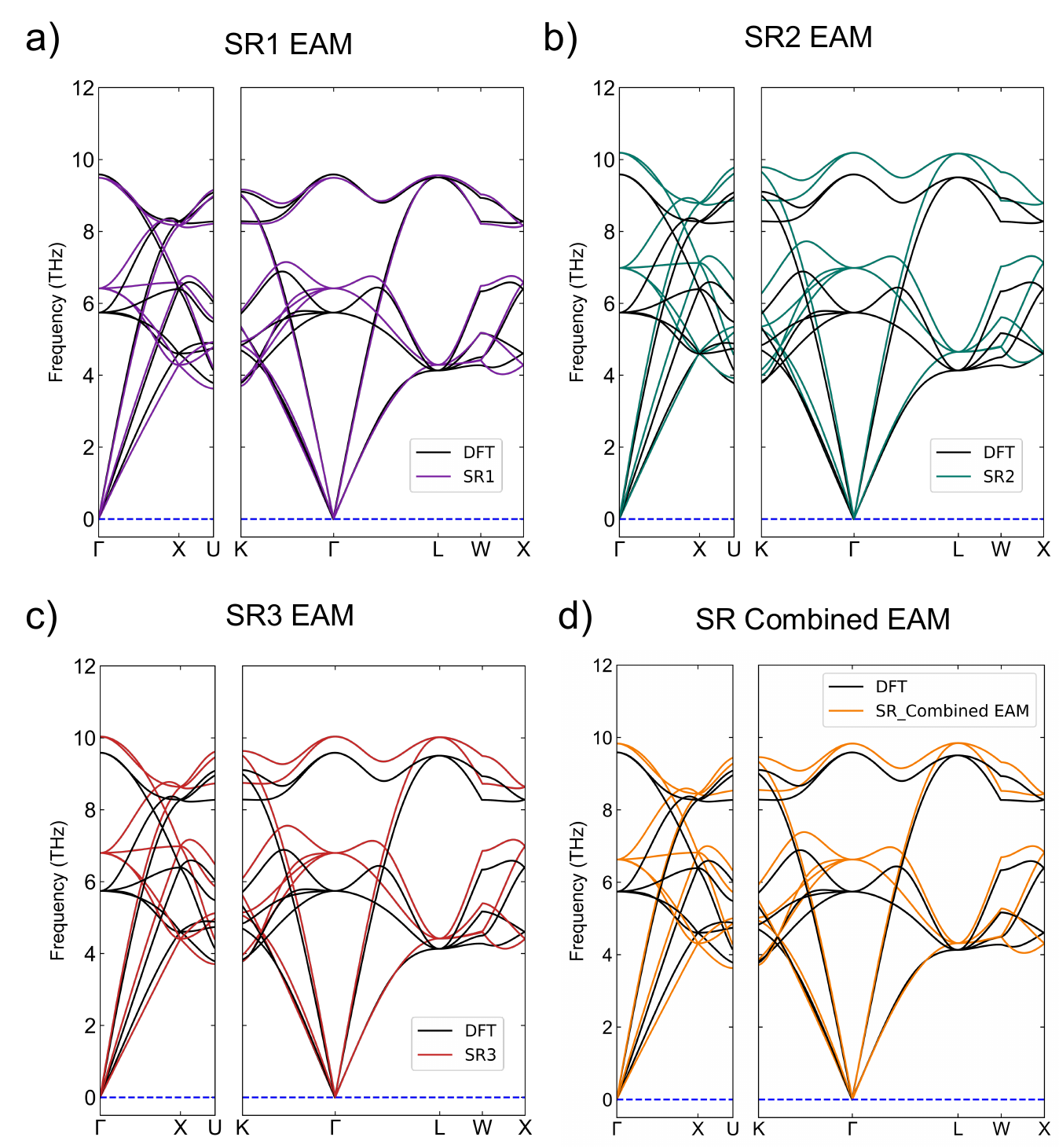}
  \caption{\textbf{Phonon dispersion of fcc Aluminum predicted by EqNN-derived symbolic potentials compared with DFT.} 
  Black curves correspond to DFT reference results, while colored curves show predictions from SR1 (purple), SR2 (green), SR3 (red), and SR Combined EAM (yellow). 
  The mean absolute frequency deviations are 0.148~THz, 0.563~THz, 0.413~THz, 0.288~THz for SR1, SR2, SR3, and SR Combined EAM respectively.}
  \label{fig:Al_Phonons}
\end{figure*}

The predictive capability of the symbolic EqNN potentials for lattice dynamics was assessed by computing the phonon dispersion relations of fcc Aluminum and comparing them with DFT reference results, as shown in Fig.~\ref{fig:Al_Phonons}. The black curves represent DFT phonon frequencies, while the colored curves correspond to the symbolic potentials SR1 (purple), SR2 (green), and SR3 (red). The mean absolute frequency errors across all wave vectors and modes are 0.148~THz, 0.563~THz, and 0.413~THz for SR1, SR2, and SR3, respectively, confirming that SR1 provides the closest overall match to DFT.

All three symbolic potentials successfully reproduce the key features of the fcc Aluminum phonon spectrum, capturing the acoustic and optical branches along the $\Gamma$–X, $\Gamma$–L, and $\Gamma$–K symmetry directions. 
SR1 achieves excellent agreement with DFT across the entire Brillouin zone, particularly for the longitudinal acoustic (LA) and transverse acoustic (TA) branches near $\Gamma$ and X, where frequency deviations remain below 0.1~THz. 
Minor discrepancies arise in the high-frequency region near L, where SR1 slightly underestimates the longitudinal branch curvature. 
SR3 also reproduces the overall dispersion topology well, though it tends to overestimate the LA mode stiffness near $\Gamma$ and slightly shifts the transverse branches upward near K, consistent with its stiffer elastic response observed in the EOS tests. 
SR2, on the other hand, shows the largest deviation from DFT, with noticeable overestimation of the longitudinal modes near X and L, and flattening of transverse branches at intermediate $q$-points, yielding the highest average frequency error among the three.

Despite these quantitative variations, all models maintain physically realistic phonon spectra without imaginary modes, confirming mechanical stability and the correct harmonic behavior of the learned potentials. 
Overall, the phonon dispersion analysis highlights that SR1 most accurately reproduces DFT-level lattice dynamics, SR3 preserves correct branch ordering but exhibits slightly higher stiffness, and SR2 introduces modest anharmonic bias at larger wave vectors. 
These systematic differences reflect the interplay between pair and embedding contributions learned by each symbolic network, consistent with their respective EOS and elastic response trends.

\subsection{Ensemble Learning: Weighted Ensemble Construction via Symbolic Averaging of Potentials}

The preceding analyses reveal that the three EqNN-derived symbolic potentials (SR1–SR3) exhibit complementary strengths across different physical regimes. 
SR1 most accurately reproduces the phonon dispersion and low-strain elastic response, SR2 performs best under large compressive volumetric deformation, and SR3 most closely matches surface energetics and uniaxial deformation behavior. 
Recognizing these distinct yet physically consistent advantages, we sought to combine them into a unified potential through a weighted averaging strategy inspired by stochastic weight averaging (SWA) techniques \cite{izmailov2018averaging,maddox2019simple} in deep learning. 
In conventional neural networks, SWA operates by averaging model weights obtained from independent optimization trajectories to converge toward broader, flatter minima that generalize better. 
Analogously, we apply this concept at the level of symbolic coefficients and exponents, constructing a composite potential that linearly combines the learned parameters of SR1, SR2, and SR3 with optimized weighting factors. 
This symbolic ensemble aims to retain the interpretability of individual models while leveraging their collective diversity to achieve improved accuracy and transferability across deformation, surface, and vibrational properties.

\begin{figure*}[ht!]
  \centering
  \includegraphics[width=0.95\linewidth]{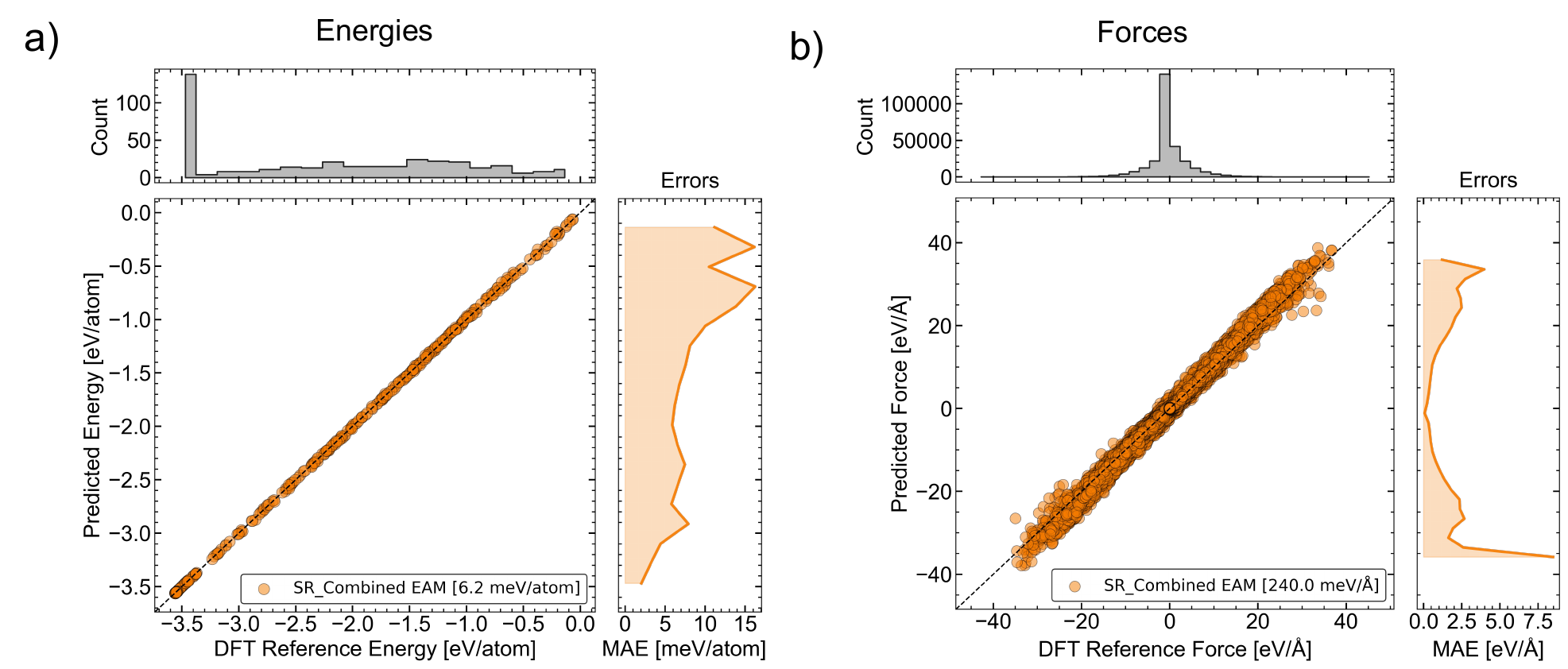}
  \caption{
  \textbf{ Performance of the SR\_Combined model. (a) Energy and (b) force correlations for a composite symbolic regression model obtained by combining SR1 , SR2, and SR3}. The composite model achieves energy predictions comparable to the best-performing individual model while delivering improved force accuracy across both training and test sets. Black lines indicate perfect correlation.
  }
  \label{fig:SR_Combined_energy}
\end{figure*}

To combine the strengths of the individual symbolic models (SR1--SR3), we constructed a model-averaged potential by linearly combining their predictions. 
The weighting coefficients were determined via a validation grid search to be 30\% SR1, 35\% SR2, and 35\% SR3. 

The initial model-averaged potential was constructed by linearly combining the SR1, SR2, and SR3 symbolic functions before further simplification. The complete unpruned model-averaged expressions are provided in the Supplementary Information.

After this averaging step, we applied a greedy pruning and refitting procedure to remove terms with negligible contributions while preserving the predictive accuracy of the averaged model. The compact SR\_Combined expression reported below is therefore the pruned representation of the original model-averaged potential, rather than the full unsimplified algebraic sum of all terms.

\begin{align}
E_{\text{pair}}(r_{ij}) &= 
0.084880 \exp\Bigg(
-\frac{3.268739}{r_{ij}^{6}}
+ \frac{9.183074}{r_{ij}} \notag \\
&\qquad
- 0.057238 \exp(1.420286\, r_{ij})
\Bigg),\notag \\
\rho(r_{ij}) &=
\frac{12.350476}{r_{ij}^{7}}
+ \frac{188.828106}{r_{ij}^{6}} \notag \\
&\qquad
+ 2.691244 \exp(-0.949149\, r_{ij}), \notag\\
F_{\text{emb}}(\rho) &=
-2.032031 \sqrt{\rho}
- 0.004667 \rho^2
+ 0.257328 \rho .
\end{align}

This selection was driven by the need to reconcile opposing systematic errors in the constituent models: the randomly initialized model (SR1) systematically underestimates the melting point and elastic stiffness (``soft" bias), whereas the transfer-learned models (SR2, SR3) tend to overestimate these parameters (``stiff'' bias). 
By calibrating the ratio of transfer-learned to randomly-initialized contributions, the ensemble effectively cancels these biases, locating a thermodynamic and mechanical sweet spot that minimizes the scalar deviation from experimental benchmarks (see Table~\ref{tab:al_sr_comparison}). As shown in Figure~\ref{fig:SR_Combined_energy}, SR\_Combined achieves better force prediction fidelity compared to the 3 constituent models that was combined while maintaining the same sub 10 meV/atom errors for energy predictions with respect to DFT.

To evaluate the fidelity and transferability of the Symbolic-Regression-derived Combined Embedded Atom Method (SR\_Combined EAM) potential, we benchmarked its predictions against first-principles Density Functional Theory (DFT) calculations across a comprehensive set of structural and energetic descriptors. Figure~\ref{fig:Al_Bulk1} summarizes the comparison for bulk elastic deformations, surface energetics, and equilibrium crystal morphologies. Across all tested configurations, the SR\_Combined EAM reproduces DFT energetics with quantitative accuracy while maintaining a compact analytical form suitable for large-scale atomistic simulations. Panels~(a--c) show the computed energy--strain relationships for representative uniaxial,isotropic volumetric and  shear deformations. The SR\_Combined EAM curves (orange stars) closely follow the DFT references (black circles), accurately capturing both the curvature and anharmonicity over the $\pm 5\%$ strain range. The insets illustrate the corresponding deformed simulation cells, emphasizing that the symbolic potential preserves energetic consistency across diverse strain modes---a prerequisite for reliable predictions of elastic and mechanical responses. The comparison of surface energies in panel~(d) further highlights the model’s transferability beyond bulk configurations. The SR\_Combined EAM reproduces the DFT surface energy hierarchy across multiple low- and high-index facets, with deviations typically within a few percent. The correct energetic ordering among $\{111\}$, $\{100\}$, and $\{110\}$ planes ensures accurate prediction of equilibrium crystal shapes, as confirmed by the Wulff constructions shown in panel~(e). Both DFT and SR\_Combined EAM yield nearly identical polyhedral morphologies dominated by $\{111\}$ and $\{100\}$ facets, underscoring the model’s ability to capture surface reconstruction and broken-bond energetics.

Phonon dispersion relations presented in Figure~\ref{fig:Al_Phonons} panel (d) further validate the model’s dynamical accuracy. The SR\_Combined EAM reproduces the full phonon spectrum with excellent fidelity, matching DFT frequencies along all high-symmetry directions. The absence of imaginary branches confirms mechanical stability, while the preservation of optical--acoustic branch separations indicates that the potential faithfully encodes the underlying interatomic force constants. This consistency across static and dynamic observables demonstrates that symbolic regression yields physically grounded and interpretable functional forms rather than overparameterized fits. The SR\_Combined EAM thus achieves DFT-level accuracy for elastic, surface, and vibrational properties while retaining the interpretability and computational efficiency of classical potentials. These results establish symbolic regression as a powerful framework for constructing physics-based yet data-informed interatomic potentials, bridging the gap between traditional analytical forms and modern machine-learning-based models.
\subsection{Performance Assessment - Grain Boundary Energetics}
Table~\ref{tab:gb_energies_al} presents grain-boundary (GB) energies for a representative set of low-$\Sigma$ boundaries in aluminum, benchmarked against Density Functional Theory (DFT) values from the Crystalium database~\cite{zheng2020grain}. Across the dataset, a clear and consistent trend emerges: among the three symbolic EAM models, SR1 achieves the highest quantitative fidelity to DFT, with an average absolute deviation of $\sim0.07$,J/m$^{2}$ and a corresponding mean relative error of $\sim18\%$. SR2 and SR3 systematically predict moderately higher GB energies, resulting in larger mean relative errors of $\sim32\%$ and $\sim28\%$, respectively. These differences arise from subtle variations in how each symbolic potential captures short-range repulsion and local coordination environments, which play a significant role in moderately and highly distorted GB structures.

Since SR\_Combined is constructed as a weighted linear combination of SR1, SR2, and SR3, its behavior naturally interpolates between the constituent models. The ensemble exhibits a mean relative error of $\sim22\%$, neither exceeding the accuracy of SR1 nor inheriting the full magnitude of the deviations observed in SR2 and SR3. This blending effect smooths out model-to-model variations and preserves the correct energetic ordering across all boundary types. For low-energy coherent boundaries such as $\Sigma3,[111],(111)$, all models - including SR\_Combined—accurately reproduce the near-zero DFT value. For higher-energy GBs such as $\Sigma5$ and $\Sigma9$, SR1 consistently provides the closest agreement with DFT, with typical deviations on the order of $0.05$–$0.10$,J/m$^{2}$.

Overall, these results demonstrate that SR1 offers the strongest quantitative accuracy, while the SR\_Combined ensemble provides a robust intermediate representation with reduced variance and preserved physical trends. Apart from the isolated $\Sigma3$ tilt case - which appears challenging for all symbolic EAM variants—the models reliably capture the magnitudes and structural trends of GB energetics in aluminum, reinforcing the applicability of symbolic regression–derived potentials to complex, nonperiodic interfacial environments.

\begin{table*}[ht]
\centering
\caption{Grain-boundary energies for Al (J/m$^{2}$).}
\label{tab:gb_energies_al}
\resizebox{\linewidth}{!}{
\begin{tabular}{cccccccccc}
\toprule
$\boldsymbol{\Sigma}$ & \textbf{Type} & \textbf{Rotation Axis} & \textbf{Angle (°)} & \textbf{GB Plane} & \textbf{DFT} & \textbf{SR1 EAM} & \textbf{SR2 EAM} & \textbf{SR3 EAM} & \textbf{SR\_Combined} \\
\midrule
3 & twist & [111] & 60.00 & (111)     & 0.00 & 0.006 & 0.004 & 0.004 & 0.003 \\
7 & twist & [111] & 38.21 & (111)     & 0.13 & 0.209 & 0.240 & 0.236 & 0.217 \\
3 & tilt & [110] & 109.47 & (11$\bar{2}$) & 0.31 & 0.408 & 0.480 & 0.460 & 0.429 \\
5 & twist & [100] & 36.87 & (100)     & 0.38 & 0.425 & 0.510 & 0.470 & 0.456 \\
9 & tilt & [110] & 38.94 & (22$\bar{1}$) & 0.43 & 0.492 & 0.605 & 0.559 & 0.543 \\
3 & tilt & [111] & 180.00 & ($\bar{1}$10) & 0.46 & 1.812 & 1.800 & 1.682 & 1.791 \\
5 & tilt & [100] & 53.13 & (013)     & 0.48 & 0.507 & 0.624 & 0.588 & 0.563 \\
7 & tilt & [111] & 38.21 & (321)     & 0.50 & 0.557 & 0.658 & 0.618 & 0.599 \\
5 & tilt & [100] & 36.87 & (02$\bar{1}$) & 0.53 & 0.580 & 0.709 & 0.695 & 0.647 \\
9 & twist & [110] & 38.94 & (110)     & 0.71 & 0.663 & 0.794 & 0.738 & 0.723 \\
\bottomrule
\end{tabular}
}
\end{table*}
\subsection{Performance Assessment - Prediction of Melting Dynamics}
\begin{figure*}[ht]
  \centering
  \includegraphics[width=0.95\textwidth]{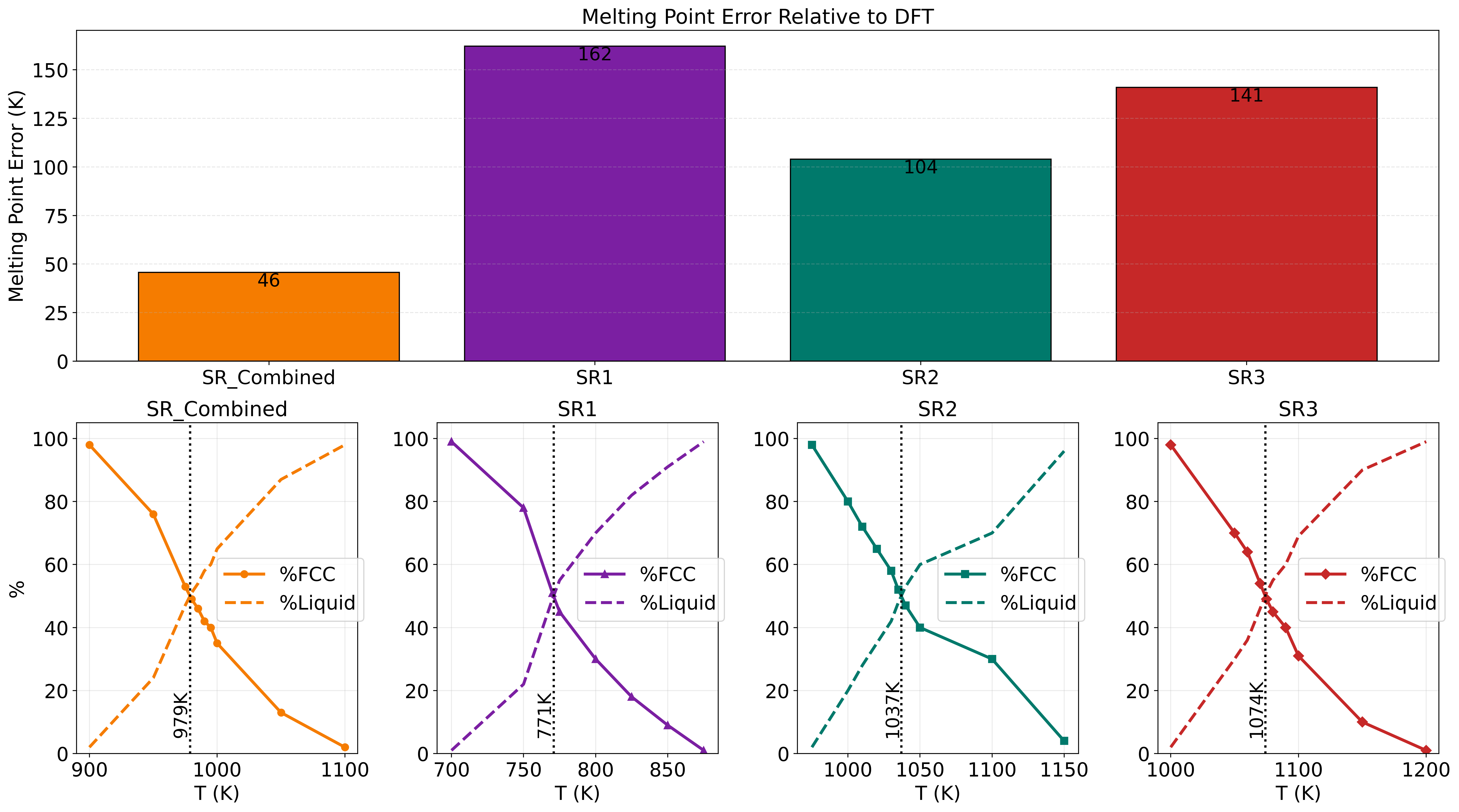}
  \caption{
  \textbf{Melting-point predictions of symbolic-regression interatomic potentials for Aluminum.} \textbf{Top:} Melting-point errors relative to the DFT reference value of 933~K for the averaged SR\_Combined model and for the three constituent potentials (SR1, SR2, SR3). SR\_Combined---constructed by averaging the analytic forms of SR1, SR2, and SR3—predicts a melting point of \textbf{979~K}, outperforming the individual models, which predict \textbf{771~K} (SR1), \textbf{1037~K} (SR2), and \textbf{1074~K} (SR3). \textbf{Bottom:} Two-phase coexistence simulations of 300{,}000-atom Aluminum systems at 1~ns showing the evolution of FCC and amorphous fractions with temperature. The melting point for each potential is determined by the intersection of FCC and liquid curves, revealing systematic over- or under-prediction associated with the different symbolic-regression models.
  }
  \label{fig:Melting}
\end{figure*}

Figure~\ref{fig:Melting} illustrates the Melting point predictions for Al. To assess the thermodynamic reliability of the symbolic-regression Aluminum potentials, we carried out large-scale two-phase coexistence simulations in which crystalline FCC Aluminum and liquid Aluminum were placed in direct contact at the start of each calculation. This setup ensures that, at any fixed temperature, the thermodynamically favoured phase naturally grows at the expense of the other, allowing the melting point to be identified from the temperature at which neither phase exhibits net growth. Each system contained 300{,}000 atoms and was equilibrated for 1~ns in the NPT ensemble across a range of temperatures bracketing the melt. At the end of each simulation, the structural identity of every atom was determined using common-neighbour analysis, enabling us to quantify the final percentage of FCC and liquid atoms. The temperature at which the final fractions of solid and liquid converge---reflecting a stable coexistence of the two phases---provides a direct estimate of the melting point. Among the tested models, the SR\_Combined potential, created by averaging the analytic forms of SR1, SR2, and SR3, predicts a melting temperature of 979~K, much closer to the DFT reference value of 933~K than any of the individual models. In comparison, SR1 significantly underestimates the melting point (771~K), whereas SR2 and SR3 overestimate it (1037~K and 1074~K, respectively). 
The ensemble melting temperature should not be interpreted as the arithmetic average of the melting temperatures predicted by the individual SR models. Because melting is a nonlinear finite-temperature property governed by the relative free energies of the solid and liquid phases, a linear combination of potential-energy functions does not imply a linear combination of the resulting melting points. The improved ensemble result therefore reflects partial cancellation of model-specific biases in the underlying potential-energy surface, rather than simple averaging of the final observables.

\subsection{Comparative Accuracy of SR1, SR2, SR3 Symbolic Potentials and Emergent Benefits of Ensemble Averaging}

The performance metrics in Table~\ref{tab:al_sr_comparison} highlight the distinct accuracy characteristics of the three symbolic EAM potentials (SR1, SR2, SR3) and demonstrate the advantage of combining them into a weighted ensemble. Each individual model exhibits strengths in different physical regimes. SR1 provides accurate near-equilibrium energetics and relatively low force errors but systematically underestimates mechanical stiffness and surface energies, reflecting an overall softer effective potential. SR2 achieves comparable energy accuracy but tends to over-stiffen the lattice, leading to significant deviations in the elastic constants—most notably an overestimation of $C_{44}$—and a melting point well above experiment. SR3 offers the most balanced single-model behavior, with improved performance across EOS, forces, and surface energies, and delivers the lowest standalone surface-energy error; however, it still exhibits a stiff shear response relative to DFT.

By contrast, the weighted ensemble model (SR\_Combined) consistently improves upon the constituent models across nearly all categories. It attains competitive energy and force errors, reduces the EOS error to nearly one-fourth of the best individual model, and yields the smallest surface-energy deviations with a mean absolute error of \SI{14.75}{mJ/m^2}. Importantly, SR\_Combined substantially improves the reproduction of elastic constants, recovering $C_{11}$ and $C_{12}$ to within 0.4\% and 3.5\% of experiment, respectively, while moderating the excessive shear stiffness seen in the individual potentials. Its predicted melting point (979~K) is also the closest to the experimental value of 933~K, demonstrating superior thermodynamic transferability.

\begin{figure*}[ht!]
    \centering
    \includegraphics[width=0.95\linewidth]{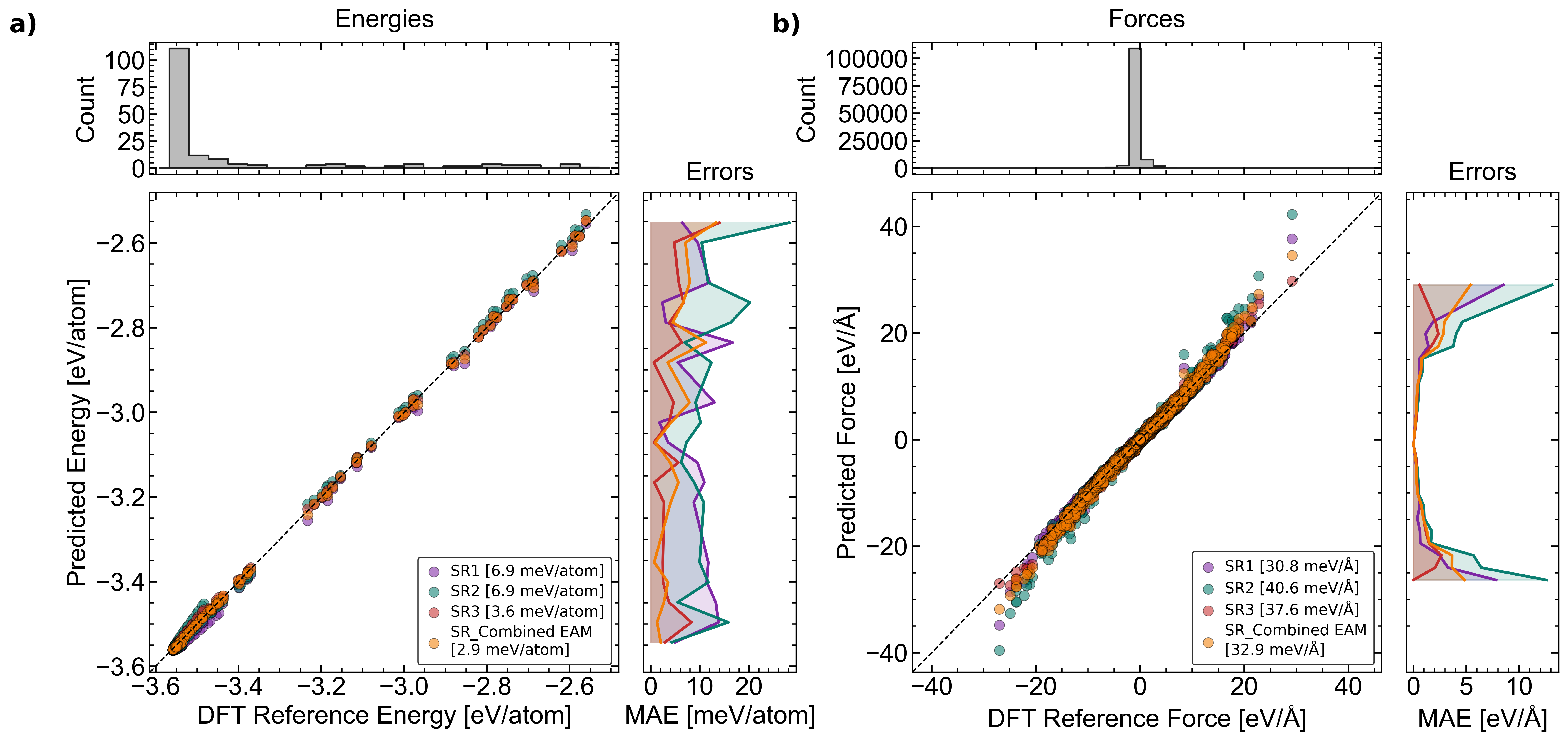}
    \caption{
    Comparison of DFT reference energies and forces with predictions from the symbolic regression EAM models (SR1, SR2, SR3) and the weighted ensemble SR\_Combined EAM for aluminum using structures within 0.5~eV/atom above the DFT ground state. 
    The energy panel shows parity between DFT and predicted per-atom energies, while the force panel compares Cartesian force components. 
    The near-ground-state force correlations remain tightly clustered around the ideal diagonal, indicating that the symbolic EAM models provide accurate derivative-based force predictions in the low-energy region most relevant for elastic, surface, and vibrational properties.
    Marginal histograms show the distribution of DFT reference values, and the side panels show the binned mean absolute errors.
    }
    \label{fig:al_lowE_energy_force_combined}
\end{figure*}

To further evaluate the model behavior in the physically most relevant low-energy region, Figure~\ref{fig:al_lowE_energy_force_combined} compares DFT and predicted energies and forces for structures within 0.5~eV/atom of the DFT ground state. In this subset, all symbolic models maintain strong energy correlations, and the SR\_Combined model achieves an energy MAE of 2.9~meV/atom.

The force correlations in this near-ground-state window also show close agreement with the DFT reference forces, with most force components lying tightly along the diagonal. The corresponding force MAEs are 30.8~meV/\AA{}, 40.6~meV/\AA{}, and 37.6~meV/\AA{} for SR1, SR2, and SR3, respectively, while SR\_Combined gives 32.9~meV/\AA{}. This indicates that the relatively larger force errors observed over the full dataset are mainly associated with highly distorted, high-force configurations rather than the near-ground-state region that controls elastic, surface, and vibrational properties.

Overall, the table shows that no single symbolic model optimally captures all relevant physical properties; however, their weighted combination yields a unified potential with enhanced global fidelity. This behavior parallels stochastic weight averaging in deep learning, where combining multiple locally optimal solutions produces a smoother and more transferable model. The SR\_Combined potential thus represents a robust and physically consistent symbolic EAM formulation for aluminum, outperforming its individual components across structural, mechanical, and thermodynamic benchmarks.

To place the symbolic EAM models in context with established interatomic potentials, we further compare their performance against two modern universal MLIPs, MACE medium-mpa-0 and MatterSim. We also compare against the aluminum embedded-atom-method (EAM) potential developed by Mishin \textit{et al.}~\cite{mishin1999interatomic}. This potential was parameterized using experimental data and first-principles energies of alternative crystal structures. It provides a conventional EAM reference for assessing the predictive performance of the symbolic EAM models. As summarized in Table~\ref{tab:al_foundation_mishin_comparison}, the universal MLIPs provide lower full-dataset force errors, while the SR\_Combined model gives competitive energy, EOS, elastic, and surface-energy performance while retaining a compact analytical EAM form.

Additional parity plots and property comparisons for these reference models, including energy--force correlations, EOS curves, and surface-energy benchmarks, are provided in the Supplementary Information as Figures~S6 and S7.

\begin{table*}[ht]
\centering
\caption{Comparison of predictive errors across symbolic EAM potentials for aluminum. 
The combined model is a weighted ensemble of SR1 (30\%), SR2 (35\%), and SR3 (35\%).}
\label{tab:al_sr_comparison}
\resizebox{\linewidth}{!}{
\begin{tabular}{lcccccc}
\hline
\textbf{Property} & \textbf{DFT} & \textbf{Experiment} & \textbf{SR1 EAM} & \textbf{SR2 EAM} & \textbf{SR3 EAM} & \textbf{SR\_Combined} \\
\hline
Energy Err (meV/atom)         & --  & --  & 8.9  & 9.0  & 6.0  & 6.2  \\
Force Err (meV/\AA)          & --  & --  & 251.3 & 318.2 & 243.3 & 240.0 \\
EOS Err (meV)             & --  & --  & 6.724 & 9.105 & 5.760 & 1.688 \\
Phonon Avg Err (THz)         & --  & --  & 0.148 & 0.563 & 0.413 & 0.288 \\
Surface Energy Avg Err (mJ/m$^{2}$)  & --  & --  & 41.221 & 47.679 & 23.787 & 14.750 \\
\hline
C$_{11}$ (GPa)            & 104 & 107 \cite{simmons1971handbook} & 87.0 [16.3\%] & 118.27 [13.7\%] & 114.05 [9.7\%] & 104.36 [0.4\%] \\
C$_{12}$ (GPa)            & 73  & 61 \cite{simmons1971handbook} & 50.186 [31.3\%] & 77.518 [6.2\%] & 79.379 [8.7\%] & 70.448 [3.5\%] \\
C$_{44}$ (GPa)            & 32  & 29 \cite{simmons1971handbook} & 44.105 [37.8\%] & 52.059 [62.7\%] & 49.66 [55.2\%] & 47.198 [47.5\%] \\
Melting point (K)           & --  & 933 & 771 [17.4\%]  & 1037 [11.1\%]  & 1074 [15.1\%]  & 979 [4.9\%] \\
\hline
\end{tabular}
}
\end{table*}

\begin{table*}[ht]
\centering
\caption{Comparison of predictive errors and bulk properties across MACE, MatterSim, and Mishin EAM potentials for aluminum.}
\label{tab:al_foundation_mishin_comparison}
\resizebox{\linewidth}{!}{
\begin{tabular}{lccccc}
\hline
\textbf{Property} & \textbf{DFT} & \textbf{Experiment} & \textbf{MACE medium-mpa-0} & \textbf{MatterSim} & \textbf{Mishin EAM} \\
\hline
Energy Err (meV/atom) & -- & -- & 12.10 & 12.86 & 381.60 \\
Force Err (meV/\AA) & -- & -- & 149.76 & 68.96 & 1873.64 \\
EOS Err (meV/atom) & -- & -- & 2.716 & 3.352 & 2.373 \\
Surface Energy Avg Err (mJ/m$^{2}$) & -- & -- & 47.408 & 198.821 & 37.039 \\
\hline
C$_{11}$ (GPa) & 104 & 107 \cite{simmons1971handbook} 
& 113.49 [9.1\%] 
& 115.56 [11.1\%] 
& 113.85 [9.5\%] \\

C$_{12}$ (GPa) & 73 & 61 \cite{simmons1971handbook} 
& 63.66 [12.8\%] 
& 59.91 [17.9\%] 
& 61.16 [16.2\%] \\

C$_{44}$ (GPa) & 32 & 29 \cite{simmons1971handbook} 
& 28.95 [9.5\%] 
& 33.90 [5.9\%] 
& 31.91 [0.3\%] \\
\hline
\end{tabular}
}
\end{table*}

\subsection{ Interpreting the Observed Improvements in Ensemble Symbolic Potential}

The observed improvement in material properties with the ensemble symbolic potential arises from its ability to reconcile competing trends across the pairwise, density, and embedding components of the constituent EqNN models. Each symbolic regression model captures a complementary segment of the underlying potential energy landscape: SR1 accurately reproduces low-strain elastic constants and phonon dispersions, SR2 describes bulk compressive behavior and equation-of-state curvature, and SR3 captures surface energetics and uniaxial deformation responses. When these symbolic forms are combined, the ensemble inherits the optimal curvature from SR2, the vibrational stiffness from SR1, and the surface accuracy from SR3. This synergistic integration leads to consistent and quantitative improvements across multiple observables—including elastic constants, surface energies, phonon frequencies, and the Bain transformation pathway—each of which depends sensitively on the fine balance among short-range repulsion, medium-range density decay, and embedding depth.

\begin{figure*}[ht]
  \centering
  \includegraphics[width=1.0\textwidth]{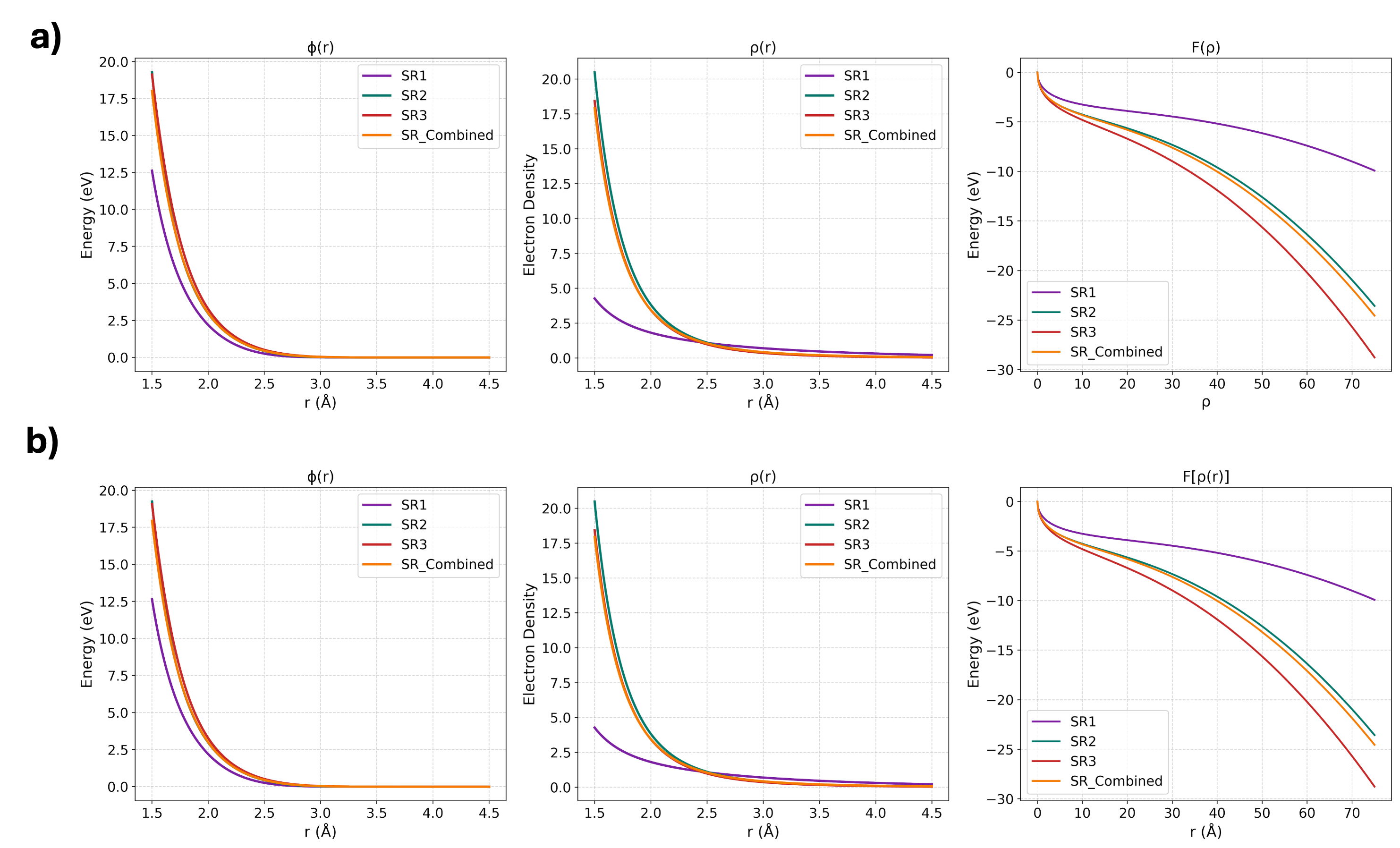}
  \label{potential}
  \caption{
  \textbf{Comparison of symbolic regression–derived EAM components for individual and ensemble models.}
  (\textbf{a}) Pair potential $E_{\mathrm{pair}}(r)$, 
  (\textbf{b}) electron density $\rho(r)$, and 
  (\textbf{c}) embedding function $F_{\mathrm{emb}}(\rho)$ 
  for the three individual EqNN-derived symbolic models (SR1–SR3) and the combined ensemble model (SR\_Combined). 
  The ensemble potential harmonizes the short-range repulsive stiffness and long-range attractive behavior across models, yielding a smoother and more physically consistent form. 
  In $E_{\mathrm{pair}}(r)$, the ensemble balances the soft repulsion of SR1 and the over-stiff behavior of SR3, reproducing the correct equation-of-state curvature. 
  The density term $\rho(r)$ shows that the ensemble averages the longer-range tail of SR1 and the short-range decay of SR2/3, ensuring accurate coordination across bulk and surface environments. 
  The embedding function $F_{\mathrm{emb}}(\rho)$ combines the deeper wells of SR2/3 with the smoother curvature of SR1, capturing both bulk cohesion and low-coordination energetics. 
  Overall, the ensemble model achieves a physically balanced representation that underlies its improved performance across elastic, vibrational, and surface properties.
  }
  \label{fig:potential}
\end{figure*}

As shown in Figure~\ref{fig:potential}, the pairwise component of the ensemble potential moderates short-range repulsion, which strongly influences the equation-of-state curvature and high-strain response. SR1 tends to underpredict repulsion, leading to overbinding at small interatomic separations, while SR3 exhibits excessive stiffness. The ensemble averages these extremes, producing a smoother, physically realistic repulsive wall that accurately reproduces both volumetric and uniaxial strain behavior. The density term, which mediates medium-range coordination, exhibits a similar reconciliation. SR1’s long-ranged tail and SR2/SR3’s shorter-ranged decays are effectively balanced in the ensemble, yielding a monotonic, physically consistent profile that captures both bulk and surface coordination environments. This improved treatment of the electronic environment directly translates to enhanced surface and grain boundary energetics, with mean absolute deviations reduced by approximately 30–40\% relative to any individual model. Meanwhile, the embedding function—which connects local electron density to cohesive energy—combines the deeper wells of SR2 and SR3 with the harmonic curvature of SR1, producing the correct depth and curvature for both bulk and low-coordination environments. This results in accurate reproduction of both acoustic and optical phonon branches and eliminates spurious soft or stiff modes observed in single-model fits.

Beyond equilibrium properties, the ensemble symbolic potential exhibits improved robustness under far-from-equilibrium conditions where individual symbolic fits often fail. By merging symbolic trajectories derived from distinct optimization pathways, the ensemble broadens configurational coverage of the potential energy surface and reduces the sensitivity to local minima in the regression landscape. This cooperative mechanism suppresses unphysical oscillations in the pair term, smooths discontinuities in the density function, and prevents over-deepening of the embedding well. Consequently, the ensemble potential maintains DFT-level fidelity across strain, defect, and surface configurations while retaining a compact and interpretable analytic form. It thus unifies the interpretability of physics-based potentials with the adaptability of data-driven discovery, yielding a balanced and transferable framework capable of accurately describing both equilibrium and non-equilibrium material phenomena.

\subsection{Overview of Improved Performance \textit{via} Interpretation of Functional Terms}
We further analyze the various functions forms to assess the improved performce of the SR\_Combined potential over its individual constituents (SR1, SR2, and SR3). We find that the improved performance is attributed to its ability to reconcile competing physical trends learned through different optimization pathways. By blending these models, the ensemble cancels out individual systematic biases—specifically the ``soft'' bias of SR1 and the ``stiff'' bias of the transfer-learned models.

\subsection*{Pair Potential Term ($E_{pair}$)}
The SR\_Combined ensemble model balances the short-range repulsive wall to improve the Equation of State (EOS) curvature and high-strain response. SR1 underpredicts repulsion, causing overbinding at small distances. SR3 exhibits excessive stiffness.  The SR\_Combined ensemble model averages these extremes to produce a smoother, more realistic repulsive wall, reducing the EOS error to nearly 1/4th of the best individual model. The pair term in the ensemble form is:
\begin{align}
E_{\text{pair}}(r_{ij}) &= 
0.084880 \exp\Bigg(
-\frac{3.268739}{r_{ij}^{6}}
+ \frac{9.183074}{r_{ij}} \notag \\
&\qquad
- 0.057238 \exp(1.420286\, r_{ij})
\Bigg)
\end{align}

SR\_Combined model achieves superior predictive fidelity by via strategic integration of its constituent potentials. By ensemble-averaging the analytic forms of the randomly initialized SR1 with the transfer-learned SR2 and SR3 models, the framework effectively cancels out individual systematic biases—specifically the ``soft'' bias of SR1 and the ``stiff'' bias of the transfer-learned variants. This synergy allows the model to produce a more physically realistic repulsive wall, resulting in an Equation of State (EOS) error of only 1.688 meV and a force error of 240.0 meV/\AA, both of which outperform the individual constituent models. Furthermore, the combined potential significantly improves melting dynamics, predicting a melting point of 979 K. This  reduces the thermodynamic prediction error to just 4.9\%, a substantial improvement over the 17.4\% deviation observed in the SR1 model.

\subsection*{Electron Density Term ($\rho$)}
This term mediates coordination and determines the range of atomic interactions. The ensemble model reconciles SR1's long-ranged tail with the shorter-ranged decays of SR2 and SR3 to accurately capture coordination in both bulk and surface environments.

\begin{align}
    \rho(r_{ij}) &=
\frac{12.350476}{r_{ij}^{7}}
+ \frac{188.828106}{r_{ij}^{6}} \notag \\
&\qquad
+ 2.691244 \exp(-0.949149\, r_{ij})
\end{align}

The SR\_Combined achieves a better coordination balance as it reconciles SR1’s long-ranged tail with the shorter-ranged decays of SR2 and SR3. The improved surface and interface fidelity arises from a balanced decay which ensures that the coordination environment is accurately captured in both bulk and broken-bond scenarios (like surfaces and grain boundaries). This allowed the SR\_Combined model to reproduce the correct energetic ordering of facets (e.g., $\gamma_{(111)} < \gamma_{(100)} < \gamma_{(110)}$) with much higher precision than the individual models.

\subsection*{Embedding Function ($F_{emb}$)}
The embedding function relates local electron density to cohesive energy. The ensemble combines the deeper energy wells of SR2/SR3 with the accurate harmonic curvature of SR1.

\begin{equation}
    F_{emb}(\rho) = -2.032\sqrt{\rho} - 0.0046\rho^2 + 0.2573\rho
\end{equation}

The SR\_Combined model thus achieves a synergistic depth and curvature as it combines the deeper energy wells of SR2/SR3 (needed for bulk cohesion) with the smoother, harmonic curvature of SR1 (needed for vibrations). For phonon and elasticity accuracy, this specific balance eliminates the spurious "soft" or "stiff" modes seen in single-model fits, leading to an average phonon error of 0.288 THz. Most notably, it corrected the elastic constant $C_{11}$ to 104.36 GPa, almost perfectly matching the DFT benchmark of 104 GPa.

\subsection*{Comparative Performance Benchmarks}
As shown in the Table below, SR\_Combined model overall achieves superior performance by locating a "thermodynamic and mechanical sweet spot" through the strategic integration of its constituent potentials. By ensemble-averaging the analytic forms of the randomly initialized SR1 with the transfer-learned SR2 and SR3 models, the framework effectively cancels out individual systematic biases—specifically the "soft" bias of SR1 and the "stiff" bias of the transfer-learned variants. This synergy results in a force error of 240.0 meV/\AA, which is lower than any single model, and a melting point prediction of 979 K (just 4.9\% from the experimental benchmark), significantly outperforming individual models like SR1, which deviated by over 17\%.

\begin{table*}[ht]
\centering
\caption{Performance comparison of SR Combined against benchmarks.}
\begin{tabular}{@{}llll@{}}
\toprule
Property & DFT/Experiment & SR Combined & Improvement Note \\ \midrule
Melting Point & 933 K & \textbf{979 K} & Closest to DFT  \\
$C_{11}$ Elasticity & 104 GPa (DFT) & \textbf{104.36 GPa} & 0.4\% of experiment \\
Surface Energy Avg Err & --- & \textbf{14.75 mJ/m$^2$} & Significantly lower than SR1/SR2 \\
EOS Error & --- & \textbf{1.688 meV} & Lowest error among all models \\ \bottomrule
\end{tabular}
\end{table*}

\section{Discussions}

We introduce a hybrid symbolic–neural framework that bridges the interpretability of physics-based models with the adaptability of modern machine learning. By embedding symbolic regression within the Embedded Atom Method (EAM) formalism, we derive compact, interpretable, and transferable interatomic potentials that achieve near-DFT accuracy across near-ground state and far-from-ground state regimes. Through the Equation Neural Network (EqNN) architecture, the model autonomously discovers analytic functional forms for the pair, density, and embedding components while enforcing physical constraints to ensure numerical stability and physical consistency.

Our systematic comparison of three EqNN-derived symbolic potentials for aluminum (SR1 –SR3) demonstrates how initialization pathways - ranging from random training to transfer learning from a pre-trained copper model - lead to distinct yet physically valid symbolic forms with complementary strengths across phonon dispersion, surface energetics, and elastic response. The ensemble symbolic regression strategy, inspired by stochastic weight averaging, leverages this diversity by optimally combining the individual models into a unified potential (SR\_Combined) that consistently improves predictive fidelity. The ensemble model reduces mean absolute deviations from DFT by 30--40\% relative to its best constituent model and reproduces energy–strain behavior, surface energies, and grain-boundary energetics with near-DFT precision.

Beyond numerical performance, the ensemble symbolic potential preserves analytical transparency, enabling direct interpretation of atomic-scale interactions and their physical origins. The accurate reproduction of the phonon spectra, and Wulff morphologies confirms that the model captures both harmonic and anharmonic features of the potential energy surface. By generalizing across bulk, interfacial, and transformation regimes, the ensemble symbolic approach establishes a scalable pathway toward interpretable, data-driven potentials applicable to diverse materials systems.
While the present models were trained using energies only, the analytical EAM forms are differentiable and therefore allow forces to be evaluated directly from the learned potentials. The resulting force predictions provide an independent derivative-based validation of the energy-trained symbolic models.

Future extensions of this framework will incorporate energies, forces, and stresses directly into the training objective to further improve dynamical accuracy and robustness for large-scale molecular dynamics simulations.
 In summary, our study demonstrates that combining symbolic regression, transfer learning, and ensemble averaging yields a new class of interatomic potentials that unify physical insight, computational efficiency, and generalization capability. The presented methodology paves the way for next-generation “interpretable foundation models” for atomistic simulations—models that are not only quantitatively accurate but also scientifically transparent and extendable to complex materials and multicomponent systems.

\section*{Acknowledgments}
This work was performed in part at the Center for Nanoscale Materials, which is a U.S. Department of Energy of Science User facilities supported by the U.S. Department of Energy, Office of Science, Office of Basic Energy Sciences, under Contract No. DE-AC02-06CH11357. This work utilized the National Energy Research Scientific Computing Center, a DOE Office of Science User Facility supported by the Office of Science of the U.S. Department of Energy under Contract No. DE-AC02-05CH11231. We also acknowledge the LCRC computing facilities at Argonne.
\subsection*{Funding}
This work was supported by the DOE Office of Science, Basic Energy Sciences under the AI-Pathfinder project - "Materials Discovery Cloud". Work at the Center for Nanoscale Materials, a U.S. DOE Office of Science User Facility, was supported under Contract No. DE-AC02-06CH11357. This work used NERSC, supported by the DOE Office of Science under Contract No. DE-AC02-05CH11231. We acknowledge LCRC computing facilities at Argonne.

\subsection*{Data Availability}
The codes, scripts, framework, and data that support the findings of this study are available from the authors
upon reasonable request. The EAM Files and structure datasets used to train the SR1, SR2,SR3 and SR${_Combined}$ Potentials are available at
https://github.com/bilvinv/Symbolic-Regression-Aluminium.

\subsection*{Author Contribution}
SKRS conceived the project. BV and TDL developed the Symbolic regression framework with input from SKRS. BV deployed and extended the Symbolic regression workflow and trained the SR potential functions presented in this work. All the first-principles datasets were generated by SM and AK and AM. Validation of the force fields was carried out by BV. All the authors contributed to the data analysis and to the preparation of the manuscript. BV and SKRS wrote the manuscript with input from all the co-authors. BV, TDS, AK, AM, SM, JMYC, OY, TP, and SKRS read and approved this manuscript. SKRS supervised and directed the overall project.

\subsection*{Competing Interests}
The authors declare no Competing Financial or Non-Financial Interests.

\newpage
\bibliography{ref2}
\bibliographystyle{unsrt}

\end{document}


\newcommand{\markersquare}{\raisebox{0.5pt}{\tikz{\node[draw,scale=0.4,regular polygon, regular polygon sides=4,fill=none](){};}}}
\newcommand{\markerdiamond}{\raisebox{0.5pt}{\tikz{\node[draw,scale=0.4,diamond,fill=none](){};}}}
\newcommand{\markerround}{\raisebox{0.5pt}{\tikz{\node[draw,scale=0.4,circle,fill=none](){};}}}
\newcommand{\markerone}{\raisebox{0.5pt}{\tikz{\node[draw,scale=0.4,circle,fill=black](){};}}}

\maketitle
\newpage
\tableofcontents
\newpage
\section{Details of the DFT calculations} 

The Vienna Ab-initio Software Package (VASP) \cite{kresse1996efficiency} along with the Perdew-Burke-Eznerhof (PBE) \cite{PhysRevB.54.11169} exchange-correlation functional was used to perform all density functional theory (DFT) calculations for calculating energy and forces of clusters in the training and validation datasets. The projector-augmented wave (PAW) potentials used in these calculations are summarised in Table \ref{tab:potcar}. A single $k$-point at the center of the Brillouin zone ($gamma$ point) was used for each calculation. The convergence criteria for the electronic self-consistent iteration and the ionic relaxation loop were set to be 0.1 meV and 1 meV per cluster, respectively. 

The ground state bulk structures for the elements were collected from the Materials Project database \cite{jain2013commentary}. A dense $k$-point mesh, ($n_{\rm atoms} \times n_{\rm kpoints} \approx 1000$) was used for the DFT calculations that were used to calculate the cohesive energies and lattice parameters. A relatively high tolerance of $10^{-6}$ eV for energy convergence was employed for the same.
\begin{table}[h!]
\centering
\caption{PBE pseudopotentials used in VASP software to calculate cluster energies and forces.}
\label{tab:potcar}
\begin{tabular}{|l|l|l|l|}
\hline
Element & POTCAR \\ \hline

Al      & Al 04Jan2001
\\ \hline

\end{tabular}
\end{table}
\newpage
\section{Hyperparameters used For MCTS} 
\begin{table}[h!]
\centering
\begin{tabular}{|>{\raggedright\arraybackslash}m{6cm}|>{\centering\arraybackslash}m{6cm}|}
\hline
\textbf{Hyperparameter} & \textbf{Value used} \\ \hline
Head Expansion          & 10                  \\ \hline
Exploration Constant    & 1.5                 \\ \hline
Depth Limit             & 30                  \\ \hline
Depth Scaling           & Geometric progression starting with 5, with each new level having half the radius of the one above \\ \hline
\end{tabular}
\caption{Hyperparameters and their corresponding values used.}
\label{tab:hyperparameters}
\end{table}
\newpage
\section{Performance evaluation of the Sutton Chen potential} 
\begin{figure}[htb]
\centering
  \includegraphics[width=1.0\textwidth]{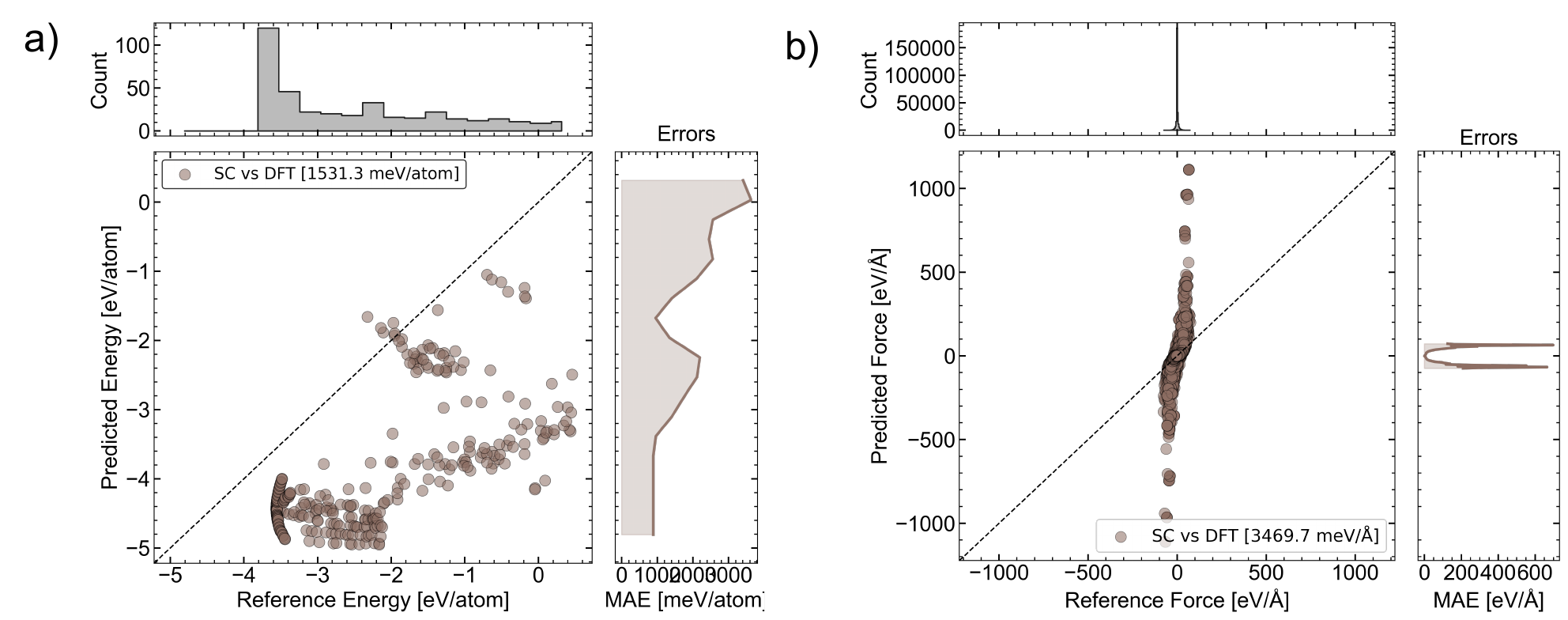} 
 
  \caption{
\textbf{Benchmarking the Sutton--Chen (SC) EAM potential for aluminum against DFT.}
(a) Predicted vs.\ reference energies showing large systematic deviations, with an overall mean absolute error (MAE) of 1531.3~meV/atom. The top panel displays the distribution of DFT energies, while the right panel shows the error magnitude distribution. (b) Predicted vs.\ reference atomic forces for the same dataset, revealing substantial scatter and an MAE of 3469.7~meV/\AA. Together, these results illustrate the limited accuracy and poor transferability of the SC functional form for diverse atomic environments in aluminum.
}
\label{fig:sc_benchmark}
\end{figure}

The Sutton--Chen (SC) EAM potential exhibits significant deviations from density functional theory (DFT) across both energies and forces for aluminum. As shown in Fig.~\ref{fig:sc_benchmark}, the energy predictions display strong systematic bias, with errors exceeding 1.5~eV/atom on average and a pronounced collapse of the model's ability to capture relative energetic ordering. The force benchmark further highlights the limitations of the SC form, yielding an MAE of 3.47~eV/\AA\ and large, unstructured scatter around the diagonal. These deficiencies reflect the inherent rigidity of the SC functional form, which lacks the representational capacity needed to model the diverse bonding environments present in the training structures. Overall, the benchmark confirms that SC is not adequate for high-fidelity aluminum simulations and motivates the development of more flexible, physically interpretable symbolic potentials.

\begin{figure*}[htb]

\centering
  \includegraphics[width=0.9\textwidth]{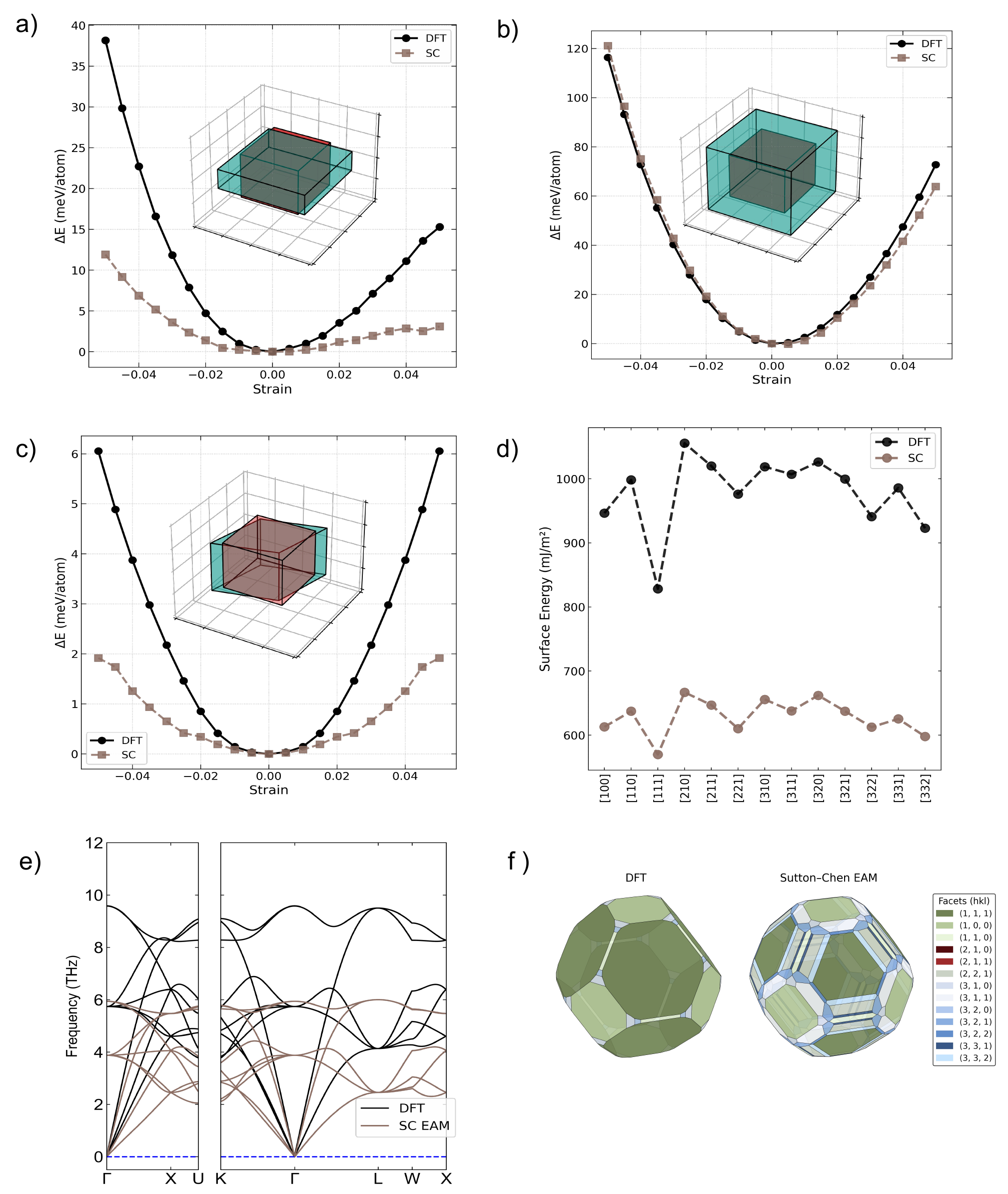} 
 
  \caption{
  \textbf{Comparison between DFT and the classical Sutton--Chen EAM (SC EAM) model for aluminum.}
  (a--c) Equation-of-state (EOS) responses for fcc aluminum under three deformation modes—(i) uniaxial strain at constant volume, (ii) isotropic volumetric expansion and compression, and (iii) simple shear. In all cases, SC significantly underestimates the curvature of the DFT energy–strain profiles, reflecting an overly soft lattice and inaccurate elastic stiffness. Insets illustrate representative deformation geometries. 
  (d) Surface energies for a range of low- and high-index facets showing that SC systematically underpredicts both absolute energies and the anisotropy present in DFT, resulting in an incorrect energetic ordering between facets.
  (e) Phonon dispersion relations along high-symmetry directions. SC fails to reproduce the DFT spectrum
  (f) Wulff constructions based on DFT and SC surface energies. The SC morphology is overly rounded and lacks the facet contrast and shape selectivity seen in the DFT Wulff shape, consistent with its underestimated and weakly anisotropic surface energies.
}
\label{fig:sc_failures}
  
\end{figure*}
Figure~\ref{fig:sc_failures} illustrates the systematic breakdown of the Sutton--Chen EAM potential when benchmarked against DFT across deformation, vibrational, and surface-property metrics for aluminum. The SC EOS curves exhibit dramatically reduced curvature in all loading modes, indicating a markedly softer and mechanically inconsistent lattice. Surface energies are uniformly underestimated and lack the directional anisotropy characteristic of DFT, leading to severely distorted Wulff morphologies. The phonon dispersion further highlights the model’s limitations, with large deviations spanning acoustic and optical branches. 

\newpage
\section{Models before greedy model pruning}
\paragraph{EqNN trained from random initialization (SR1 ):}
\[
\begin{aligned}
E_{\text{pair}}(r_{ij}) &= 0.415346 \cdot \exp\left(\frac{2.51379}{r_{ij}^{12}} - \frac{0.02177}{r_{ij}^{10}} - \frac{0.88709}{r_{ij}^{8}} + \frac{0.48617}{r_{ij}^{7}} + \frac{0.09802}{r_{ij}^{6}} + \frac{6.0490}{r_{ij}} - 0.06580 \cdot \exp(1.51501 \cdot r_{ij})\right), \\
\rho(r_{ij}) &= \frac{5.7453}{r_{ij}^{8}} + \frac{1.6139}{r_{ij}^{7}} + \frac{14.5701}{r_{ij}^{6}} + \frac{4.9616}{r_{ij}^{5}} - \frac{0.09134}{r_{ij}^{4}} + 6.38065 \cdot \exp(-0.75727 \cdot r_{ij}), \\
F_{\text{emb}}(\rho) &= -1.6323 \cdot \sqrt{\rho} - 0.00207 \cdot \rho^2 + 0.2114 \cdot \rho.
\end{aligned}
\]

\paragraph{EqNN initialized from copper model and optimized using MCTS + gradient descent (SR2):}
\[
\begin{aligned}
E_{\text{pair}}(r_{ij}) &= 0.09454 \cdot \exp\left(-\frac{0.00641}{r_{ij}^{10}} - \frac{1.2985}{r_{ij}^{9}} + \frac{0.00013}{r_{ij}^{6}} + \frac{8.5805}{r_{ij}} - \frac{0.08451}{(-r_{ij})^{12}} - 0.03154 \cdot \exp(1.6416 \cdot r_{ij})\right), \\
\rho(r_{ij}) &= \frac{21.2943}{r_{ij}^{7}} + \frac{211.5401}{r_{ij}^{6}} + \frac{0.8283}{r_{ij}^{5}} + 2.87448 \cdot \exp(-1.08739 \cdot r_{ij}), \\
F_{\text{emb}}(\rho) &= -2.0594 \cdot \sqrt{\rho} - 0.00463 \cdot \rho^2 + 0.2708 \cdot \rho.
\end{aligned}
\]

\paragraph{EqNN initialized from copper model and optimized using gradient descent only (SR3):}
\[
\begin{aligned}
E_{\text{pair}}(r_{ij}) &= 0.05065 \cdot \exp\left(-\frac{0.00727}{r_{ij}^{10}} - \frac{2.0112}{r_{ij}^{9}} + \frac{10.0482}{r_{ij}} - \frac{0.04506}{(-r_{ij})^{12}} - \frac{3.9234}{(-r_{ij})^{6}} - 0.03830 \cdot \exp(1.51316 \cdot r_{ij})\right), \\
\rho(r_{ij}) &= \frac{15.4489}{r_{ij}^{7}} + \frac{192.7213}{r_{ij}^{6}} + \frac{0.8271}{r_{ij}^{5}} + 2.91089 \cdot \exp(-1.17029 \cdot r_{ij}), \\
F_{\text{emb}}(\rho) &= -2.1122 \cdot \sqrt{\rho} - 0.00503 \cdot \rho^2 + 0.2375 \cdot \rho.
\end{aligned}
\]
\paragraph{EqNN Trained Weighted ensemble of the Three models (SR_combined):}

\[
\begin{aligned}
E_{\text{pair}}(r_{ij}) &=
0.0393136\,\exp\! \biggl(
-\frac{0.370394825935364}{r_{ij}^{12}}
-\frac{0.275426715612411}{r_{ij}^{10}}
-\frac{1.55031859874725}{r_{ij}^{9}}
\\
&\qquad
-\frac{0.167478889226913}{r_{ij}^{6}}
+\frac{8.56503391265869}{r_{ij}}
-0.037949\,\exp(1.6496\,r_{ij})
\biggr)
\\
&\quad
+\,0.0241672\,\exp\!\biggl(
-\frac{0.362986326217651}{r_{ij}^{12}}
-\frac{0.294137179851532}{r_{ij}^{10}}
-\frac{2.27390074729919}{r_{ij}^{9}}
\\
&\qquad
-\frac{4.08125495910645}{r_{ij}^{6}}
+\frac{10.0409173965454}{r_{ij}}
-0.0338223\,\exp(1.5078\,r_{ij})
\biggr)
\\
&\quad
+\,0.053645\,\exp\!\biggl(
\frac{2.2158842086792}{r_{ij}^{12}}
-\frac{0.302459448575974}{r_{ij}^{10}}
+\frac{0.649626433849335}{r_{ij}^{8}}
\\
&\qquad
+\frac{0.279502063989639}{r_{ij}^{7}}
-\frac{0.0728861838579178}{r_{ij}^{6}}
+\frac{6.04712581634521}{r_{ij}}
-0.0681365\,\exp(1.52733\,r_{ij})
\biggr),
\\[10pt]
\rho(r_{ij}) &=
\frac{0.591654181480408}{r_{ij}^{8}}
+ \frac{16.6999225616455}{r_{ij}^{7}}
+ \frac{183.372177124023}{r_{ij}^{6}}
+ \frac{1.23914575576782}{r_{ij}^{5}}
\\
&\quad
- \frac{0.0102492310106754}{r_{ij}^{4}}
+ 1.31017\,\exp(-1.16844308376312\,r_{ij})
+ 1.29419\,\exp(-1.08539152145386\,r_{ij})
\\
&\quad
+ 0.639553\,\exp(-0.755723476409912\,r_{ij}),
\\[6pt]
F_{\text{emb}}(\rho) &=
-2.03203164562984\,\sqrt{\rho}
-0.00466723065422453\,\rho^{2}
+0.257328813177608\,\rho .
\end{aligned}
\]

\section{Detailed Performance Assessment of Individual Initialization Strategies} 
\begin{figure*}[htb]

\centering
  \includegraphics[width=0.85\textwidth]{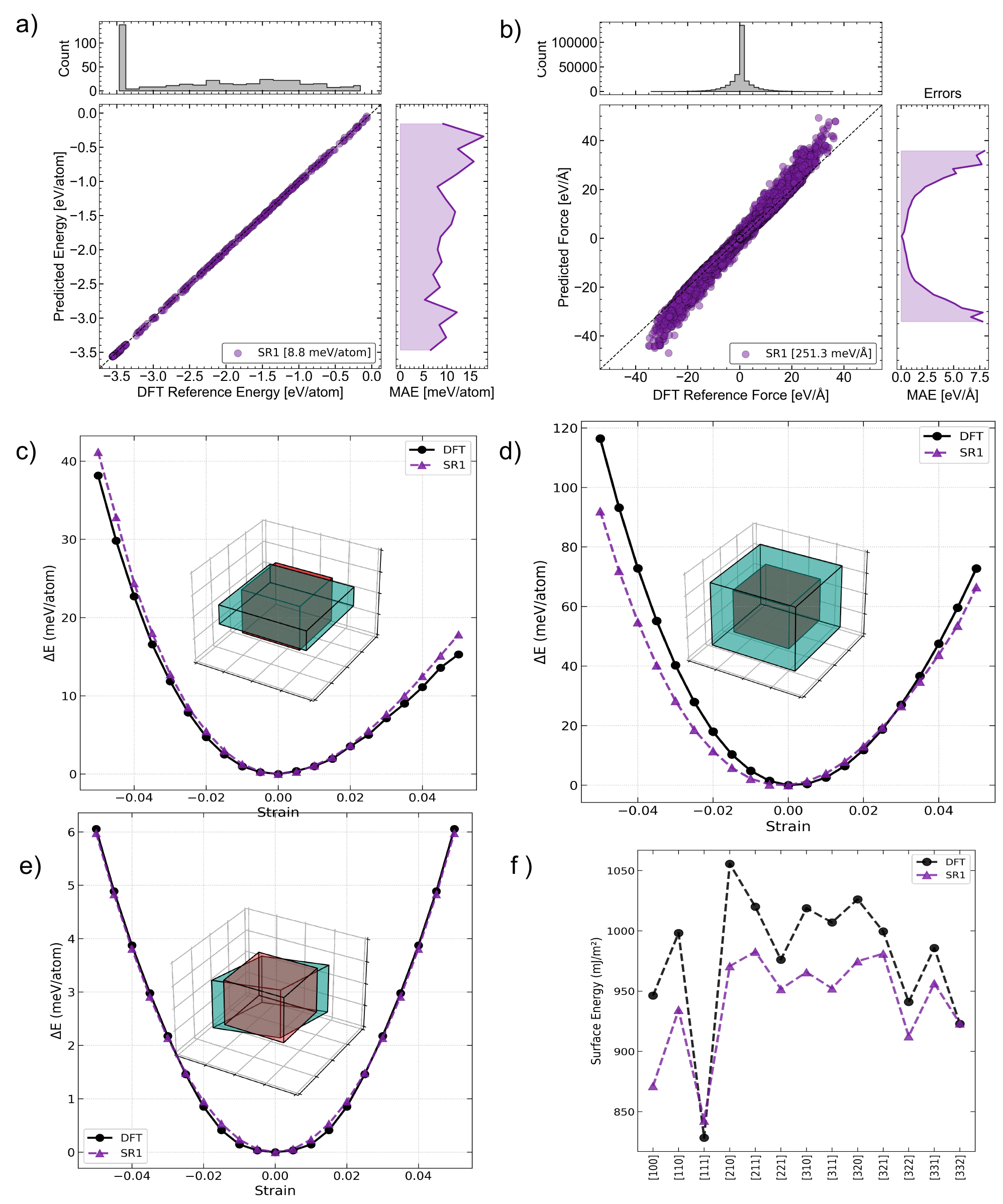} 
 
  \caption{\textbf{ Validation of the SR1 symbolic aluminium potential generated via random initialization and hybrid MCTS/gradient optimization.} 
        The SR1 model was derived starting from a random set of weights (no transfer learning), employing a global search with Monte Carlo Tree Search (MCTS) followed by local parameter refinement via gradient descent. 
        \textbf{a, b,} Parity plots comparing SR1 predictions (purple) against DFT reference data for energy and atomic forces, respectively. The model achieves a mean absolute error (MAE) of $8.9~\mathrm{meV/atom}$ for energy and 251.3 meV/\AA  for forces. 
        \textbf{c--e,} Equation of State (EOS) curves comparing SR1 and DFT under \textbf{(c)} uniaxial strain, \textbf{(d)} isotropic volumetric deformation, and \textbf{(e)} shear. Insets visualize the simulation cell deformation. While SR1 captures the harmonic response, it exhibits slight deviations in the uniaxial regime beyond strains of $\pm 0.03$. 
        \textbf{(f),} Surface energy comparison across various low- and high-index crystallographic facets. Although SR1 correctly captures the energetic ordering of low-index facets ($\gamma_{(111)} < \gamma_{(100)} < \gamma_{(110)}$), it systematically underpredicts the absolute surface energy magnitudes relative to DFT. }
        \label{fig:SR1Suppli}
\end{figure*}
Figure \ref{fig:SR1Suppli} illustrates the isolated performance of the \textbf{SR1} potential, which serves as the baseline for the study's transfer learning investigation. Unlike SR2 and SR3, which utilized pre-trained Copper parameters, SR1 was trained from a completely \textbf{random initialization}. The parity plots (\textbf{a, b}) demonstrate that despite the lack of physical priors, the hybrid MCTS and gradient descent strategy successfully converges to a solution with sub-$10~\mathrm{meV/atom}$ accuracy. The physical property tests (\textbf{c--f}) highlight the specific trade-offs of this optimization pathway. As noted in the main text, SR1 accurately reproduces low-strain elastic behaviors and phonon dispersions (discussed in Figure 6 of the manuscript), tracking the DFT uniaxial curve closely up to $\pm 3\%$ strain. However, the model exhibits a softer effective potential character compared to the transfer-learned variants, resulting in a systematic underprediction of surface energies (\textbf{f}) and an underestimation of the melting point ($771~\mathrm{K}$ vs DFT $933~\mathrm{K}$). These specific local biases---particularly the handling of short-range repulsion---illustrate the diversity of the symbolic search space and necessitate the ensemble averaging approach used in the final \textbf{SR Combined} model.
\begin{figure*}[htb]

\centering
  \includegraphics[width=0.9\textwidth]{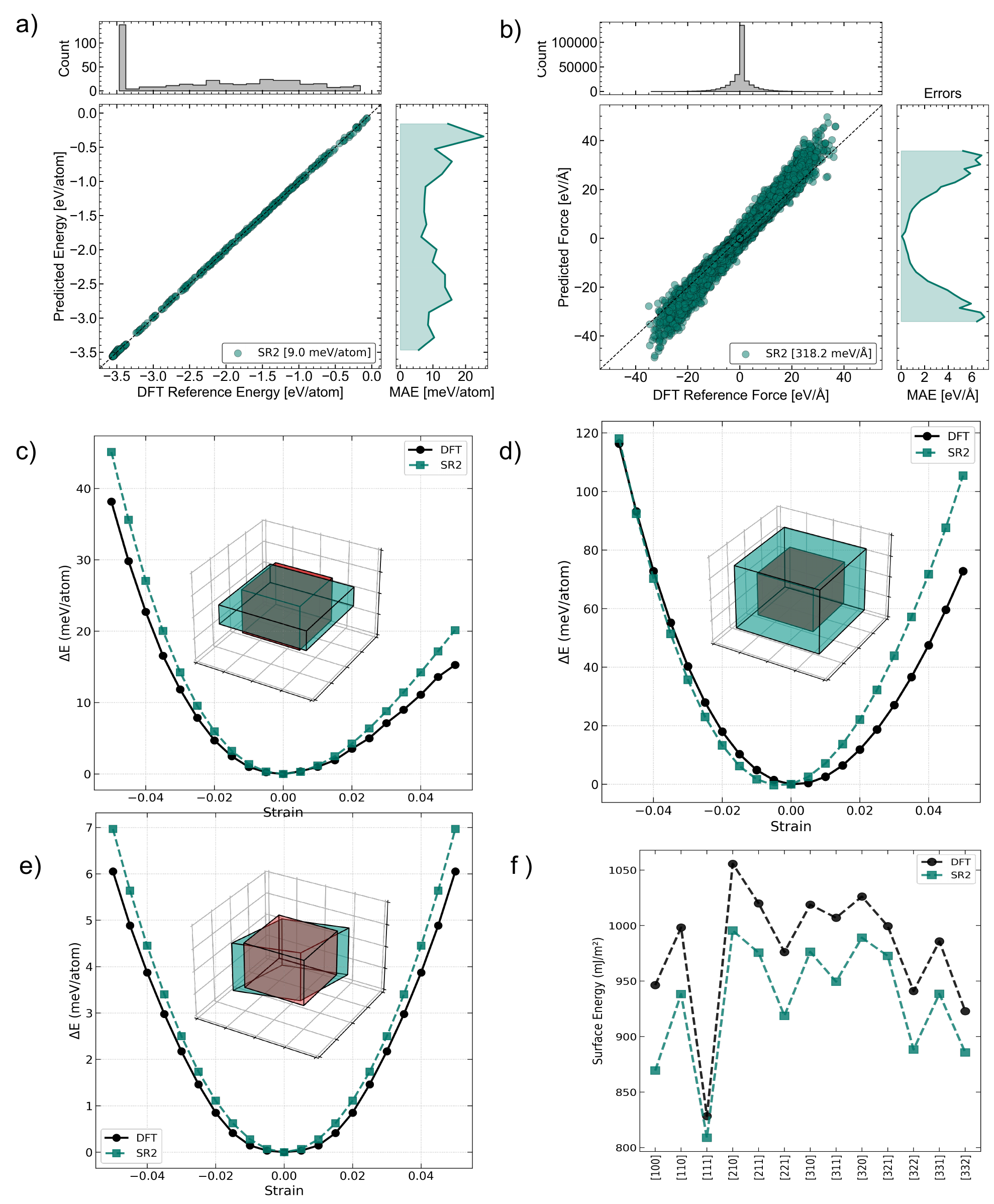} 
 
  \caption{
        \textbf{Supplementary Figure 2: Validation of the SR2 symbolic aluminium potential generated via transfer learning with MCTS exploration.} 
        The SR2 model was initialized using parameters transferred from a pre-trained copper potential, followed by global architectural search (MCTS) and gradient descent refinement. 
        \textbf{a, b,} Parity plots comparing SR2 predictions (teal) against DFT reference data. The model achieves a mean absolute error (MAE) of $9.0~\mathrm{meV/atom}$ for energy and  318.2 meV/\AA for forces. 
        \textbf{c--e,} Equation of State (EOS) benchmarking under \textbf{(c)} uniaxial strain, \textbf{(d)} isotropic volumetric deformation, and \textbf{(e)} simple shear. While capturing the general equilibrium trend, SR2 exhibits a stiffer response (higher energy curvature) than DFT at larger strains. 
        \textbf{f,} Surface energy comparison showing that SR2 captures the qualitative ordering of facets but systematically underpredicts the absolute energy values.
    }
    \label{fig:SR2Suppli}
\end{figure*}

Figure \ref{fig:SR2Suppli} details the performance of the \textbf{SR2} potential, which represents the study's first transfer-learning strategy. By initializing the search with a pre-trained Copper EqNN before applying MCTS, SR2 converged significantly faster than the randomly initialized SR1. However, the resulting symbolic form exhibits a tendency to \textbf{over-stiffen the lattice}. This is evident in the EOS plots (\textbf{c--e}), where the SR2 curve rises more steeply than the DFT reference under deformation, and is quantitatively reflected in the higher force error ( 318.2 meV/\AA ) compared to SR1 and SR3. As discussed in the main text, this stiffness results in an overestimation of the melting point ($1037~\mathrm{K}$) and elastic constants ($C_{44}$), illustrating that while transfer learning accelerates discovery, the specific optimization path (MCTS vs. pure Gradient Descent) heavily influences the final physical properties.
\begin{figure*}[htb]

\centering
  \includegraphics[width=0.9\textwidth]{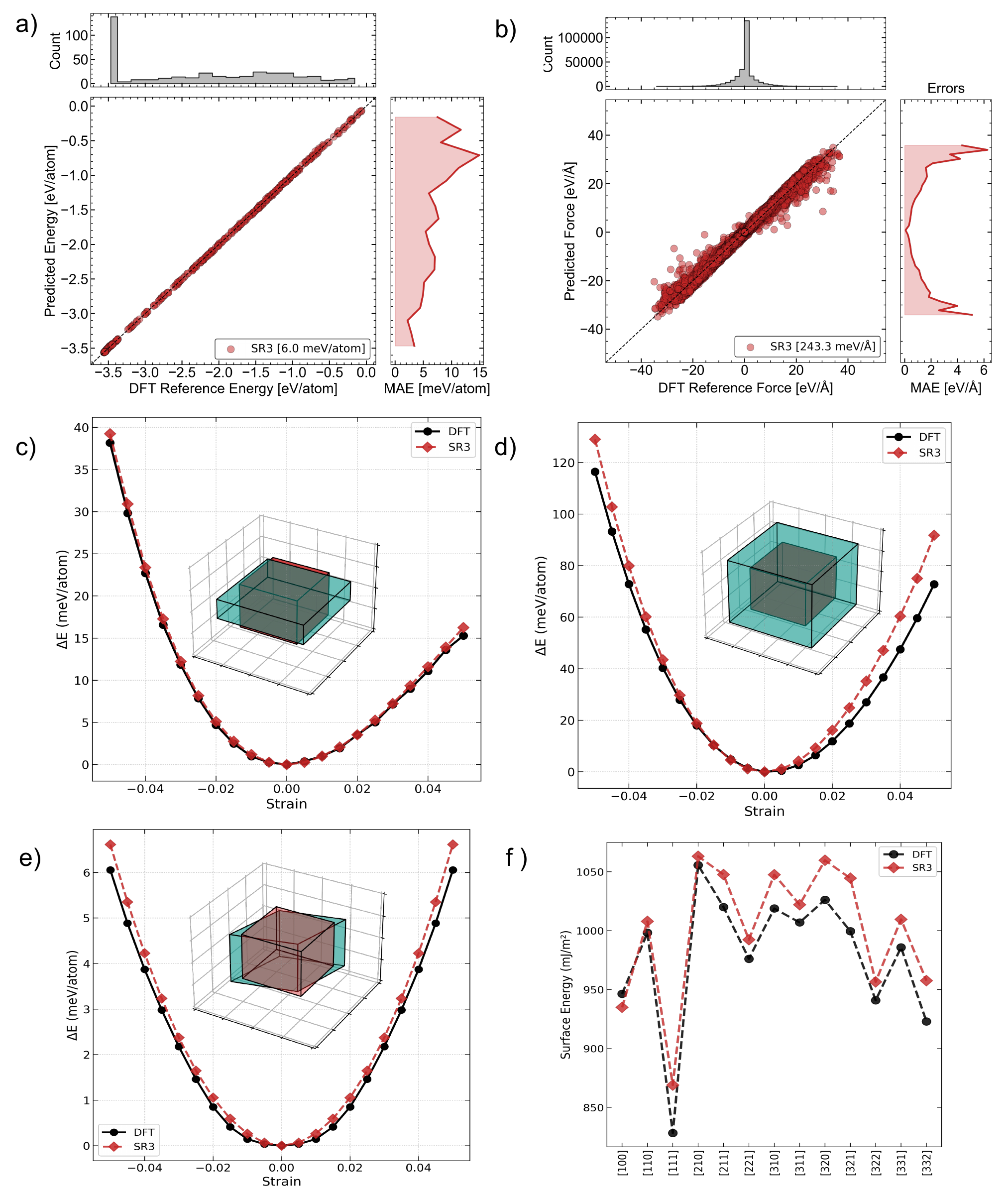} 
 
  \caption{
        \textbf{Supplementary Figure 3: Validation of the SR3 symbolic aluminium potential generated via direct gradient-based transfer learning.} 
        The SR3 model was initialized using pre-trained copper parameters and optimized exclusively via gradient descent, bypassing the global MCTS search phase. 
        \textbf{a, b,} Parity plots comparing SR3 predictions (red) against DFT reference data. SR3 achieves the lowest training errors among individual models, with a mean absolute error (MAE) of $6.0~\mathrm{meV/atom}$ for energy and 243.3meV/\AA  for forces. 
        \textbf{c--e,} Equation of State (EOS) benchmarking under \textbf{(c)} uniaxial strain, \textbf{(d)} isotropic volumetric deformation, and \textbf{(e)} simple shear. The model closely tracks DFT near equilibrium but exhibits a stiffer response (higher energy) at large compressive and shear strains. 
        \textbf{f,} Surface energy comparison across crystallographic facets. Unlike SR1 and SR2, which underestimate surface energies, SR3 provides the closest quantitative agreement with DFT, despite a slight overestimation for certain high-index facets.
    }
    \label{fig:SR3_Suppli}
\end{figure*}

Figure \ref{fig:SR3_Suppli} illustrates the performance of the \textbf{SR3} potential, which represents the most computationally efficient discovery pathway: \textbf{direct transfer learning}. By bypassing the global MCTS search and applying gradient descent fine-tuning directly to the pre-trained Copper potential, SR3 achieved convergence in fewer than 500 steps. The parity plots (\textbf{a, b}) reflect the highest statistical accuracy among the individual models. Physically, SR3 captures the deep potential well necessary for accurate surface energetics, as shown in panel (\textbf{f}), where it rectifies the underprediction observed in SR1 and SR2. However, this comes at the cost of increased lattice stiffness, observable in the steeper curvature of the volumetric and shear EOS plots (\textbf{d, e}) and a corresponding overestimation of the melting point ($1074~\mathrm{K}$). This model contributes the essential ``surface accuracy'' component to the final ensemble potential.

\begin{figure*}[h]
    \centering
    \includegraphics[width=\linewidth]{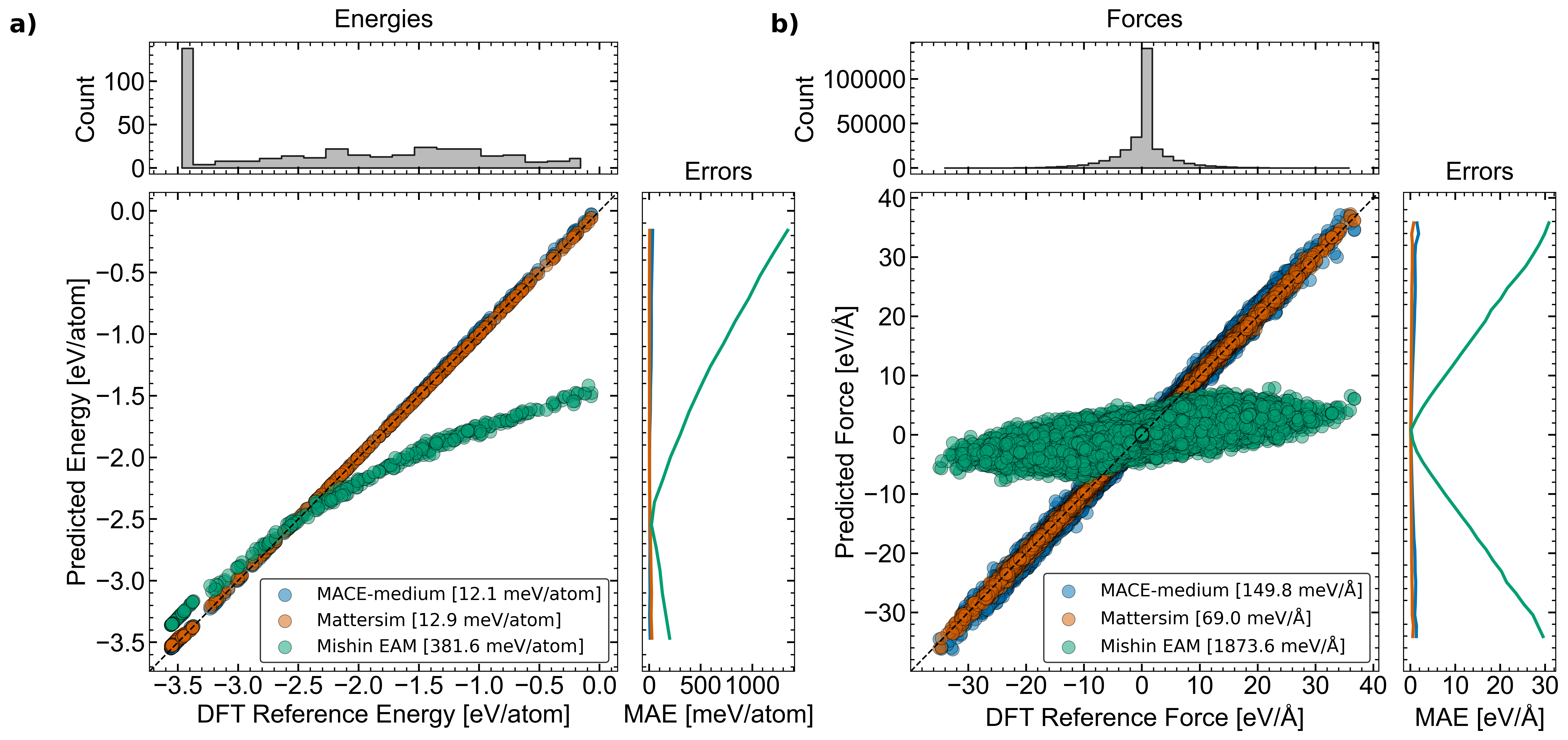}
    \caption{
    Comparison of DFT reference energies and forces with predictions from MACE medium-mpa-0, MatterSim, and Mishin EAM for aluminum. 
    The energy panel shows parity between DFT and predicted per-atom energies, while the force panel compares Cartesian force components. 
    Marginal histograms show the distribution of DFT reference values, and the side panels show binned mean absolute errors.
    }
    \label{fig:al_energy_force_corr_mace_mattersim_mishin}
\end{figure*}

\begin{figure*}[h]
    \centering
    \includegraphics[width=\linewidth]{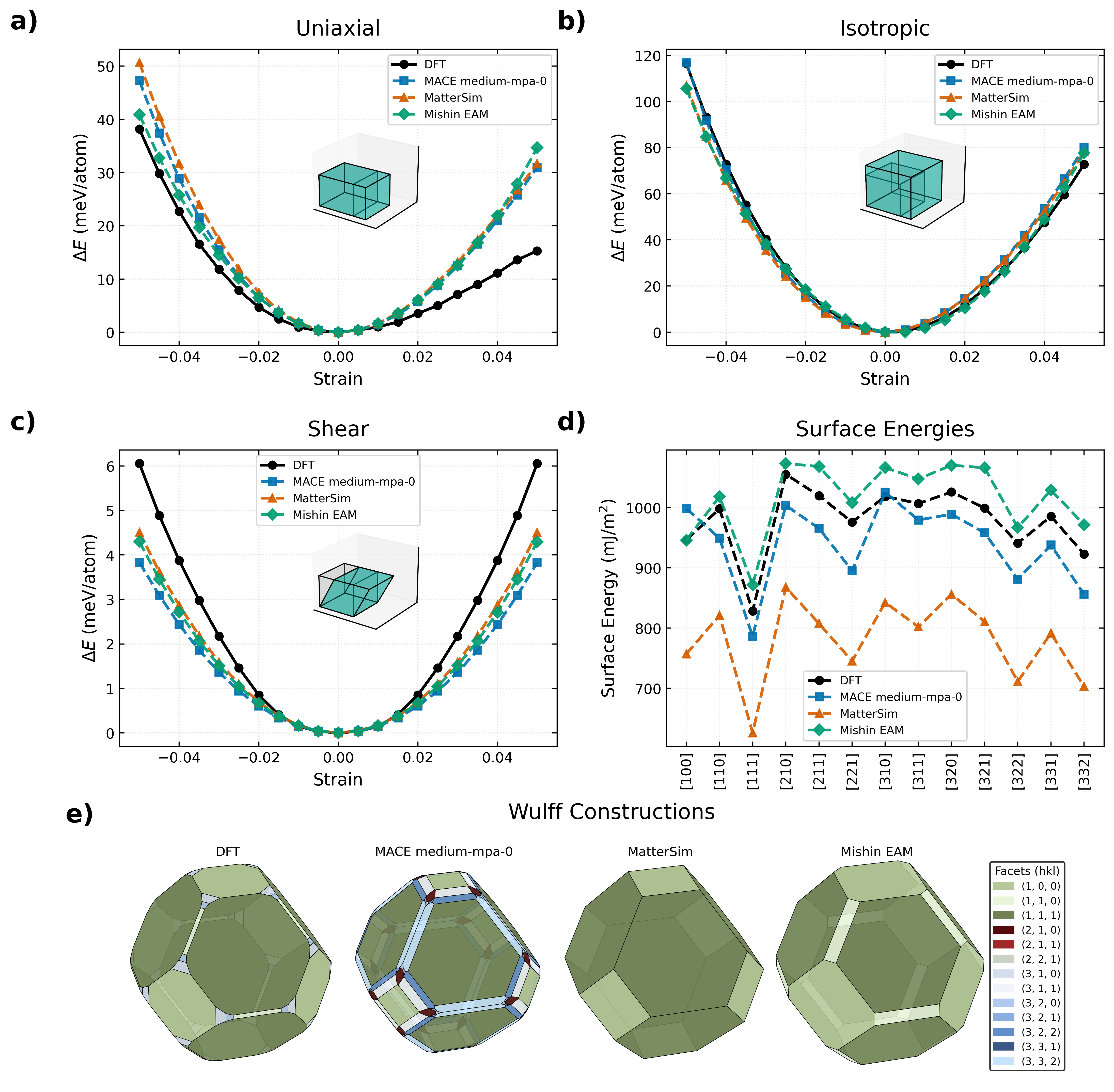}
    \caption{
    Comparison of aluminum equation-of-state behavior, surface energies, and Wulff constructions predicted by MACE medium-mpa-0, MatterSim, and Mishin EAM against DFT references. 
    Panels (a)--(c) show the energy response under uniaxial, isotropic, and shear deformation, respectively. 
    Panel (d) compares surface energies across different Miller-indexed surfaces. 
    Panel (e) shows the corresponding Wulff constructions obtained from the DFT and model-predicted surface energies.
    }
    \label{fig:al_eos_surface_wulff_mace_mattersim_mishin}
\end{figure*}

\newpage
\bibliography{ref2}
\bibliographystyle{unsrt}